\documentclass[10pt,journal,compsoc]{IEEEtran}

\usepackage{amsbsy}
\usepackage{makecell}

\usepackage{cite}
\usepackage{amsmath,amssymb,amsfonts}
\usepackage{graphicx}
\usepackage{textcomp}
\usepackage{xcolor}
\usepackage{subfigure}
\usepackage{bm}
\usepackage{bbm}

\usepackage{booktabs}
\usepackage{algorithm}
\usepackage{amsmath}
\usepackage[noend]{algpseudocode}
\usepackage{multirow}

\DeclareMathOperator*{\vertconcat}{\text{\Large $\parallel$}}
\DeclareMathOperator*{\pack}{\mathrm{\footnotesize PACK}}
\DeclareMathOperator*{\pass}{\mathrm{\footnotesize PASS}}
\DeclareMathOperator*{\fuse}{\mathrm{\footnotesize FUSE}}

\begin{document}
\title{Uniting Heterogeneity, Inductiveness, and Efficiency for Graph Representation Learning}

\author
{
 Tong~Chen, Hongzhi~Yin, Jie~Ren, Zi~Huang, Xiangliang Zhang, Hao Wang\\
   \IEEEcompsocitemizethanks{
  \IEEEcompsocthanksitem T. Chen, H. Yin, J. Ren and Z. Huang are with the School of Information Technology and Electrical Engineering, The University of Queensland.\protect\\
E-mail: tong.chen@uq.edu.au, h.yin1@uq.edu.au, thurenjie@foxmail.com, huang@itee.uq.edu.au
%\IEEEcompsocthanksitem Q. V. H. Nguyen is with the School of Information and Communication Technology, Griffith University.\protect\\
%E-mail: quocviethung.nguyen@griffith.edu.au
%\IEEEcompsocthanksitem K. Zheng is with the Big Data Research Center, University of Electronic Science and Technology of China.\protect\\
%E-mail: zhengkai@uestc.edu.cn
\IEEEcompsocthanksitem X. Zhang is with the Machine Intelligence and Knowledge Engineering Laboratory, King Abdullah University of Science and Technology.\protect\\
E-mail: xiangliang.zhang@kaust.edu.sa
\IEEEcompsocthanksitem H. Wang is with Alibaba AI Labs. E-mail: cashenry@126.com
}
\thanks{Manuscript is under review. Hongzhi Yin is the corresponding author.}
}




\IEEEtitleabstractindextext{\begin{abstract}
With the ubiquitous graph-structured data in various applications, models that can learn compact but expressive vector representations of nodes have become highly desirable. Recently, bearing the message passing paradigm, graph neural networks (GNNs) have greatly advanced the performance of node representation learning on graphs. However, a majority class of GNNs are only designed for homogeneous graphs, leading to inferior adaptivity to the more informative heterogeneous graphs with various types of nodes and edges. %These methods are commonly based on message passing via either local neighborhoods or meta paths, thus falling short when modelling sparse graphs and/or capturing important low-order contexts. 
Also, despite the necessity of inductively producing representations for completely new nodes (e.g., in streaming scenarios), few heterogeneous GNNs can bypass the transductive learning scheme where all nodes must be known during training. Furthermore, the training efficiency of most heterogeneous GNNs has been hindered by their sophisticated designs for extracting the semantics associated with each meta path or relation. In this paper, we propose \textbf{wi}de and \textbf{de}ep message passing \textbf{n}etwork (WIDEN) to cope with the aforementioned problems about heterogeneity, inductiveness, and efficiency that are rarely investigated together in graph representation learning. In WIDEN, we propose a novel inductive, meta path-free message passing scheme that packs up heterogeneous node features with their associated edges from both low- and high-order neighbor nodes. To further improve the training efficiency, we innovatively present an active downsampling strategy that drops unimportant neighbor nodes to facilitate faster information propagation. Experiments on three real-world heterogeneous graphs have further validated the efficacy of WIDEN on both transductive and inductive node representation learning, as well as the superior training efficiency against state-of-the-art baselines. 
\end{abstract}}

\maketitle

%\begin{IEEEkeywords}
%IEEE, IEEEtran, journal, \LaTeX, paper, template.
%\end{IEEEkeywords}

\IEEEpeerreviewmaketitle

\section{Introduction}

\IEEEPARstart{N}{owadays} a considerable amount of real-life applications constantly involve graph-structured data, e.g., social networks, citation records, and power grids. To fully analyse the information within large-scale graphs, learning low-dimensional representations (a.k.a. embeddings) of nodes has been a well-established solution, where high-quality node embeddings are required to retain maximal information of nodes' own features and their connectivity within the graph. With the informative yet compact node embeddings, various downstream tasks like node classification \cite{} and link prediction can be accurately performed. 

With the widely proven effectiveness of deep neural networks, their adaptation to graph-structured data has been a great success as well. Thus, graph neural networks (GNNs) have demonstrated state-of-the-art effectiveness in a wide range of graph mining applications, such as personalized recommendation \cite{wang2019neural} and molecular fingerprint mining \cite{butler2018machine}. The rationale behind GNNs is based on the message passing scheme between different nodes in a graph. For example, the graph convolutional network (GCN) \cite{kipf2017GCN} extends the convolutional operation to the non-Euclidean graph-structured data, and uses the connectivity of the graph as the filter to perform neighborhood information aggregation. At the same time, the graph attention network (GAT) \cite{velivckovic2018graph} utilizes attention mechanism to simultaneously learn the topological structure of each node's neighborhood and the distribution of node features in a graph. Despite the prosperity of GNNs, a large body of GNN models are exclusively designed for learning node representations on homogeneous graphs \cite{yun2019graph} with the assumption that a graph contains only single-typed nodes and edges. When handling heterogeneous graphs (e.g., user-item graphs in e-commerce) with various types of nodes and edges, such limitation hinders a model's capability of fully capturing the diverse semantics associated with the graph's heterogeneity. Consequently, GNNs for heterogeneous graphs are introduced in recent research \cite{wang2019heterogeneous,yun2019graph,hu2020heterogeneous,wang2019kgat}. 

Essentially, to account for the diverse semantics among different types of nodes and edges, a general remedy in heterogeneous GNNs is to utilize the notion of meta-paths \cite{sun2011pathsim,zhang2019heterogeneous,wang2019heterogeneous}, which are node sequences connected with heterogeneous edges, e.g., the \textit{author-paper-conference} meta paths in a citation graph. As a common practice, by pre-defining some specific meta path compositions, a heterogeneous graph can be decomposed into several collections of meta-paths. Then, homogeneous GNNs can thus be applied to each collection for node representation learning, such as HAN \cite{wang2019heterogeneous} and GTN \cite{yun2019graph} that respectively extend GAT and GCN to different meta paths. However, a widely acknowledged drawback of meta path-based GNNs is that, the design of meta paths are usually task-specific and requires abundant domain knowledge \cite{yun2019graph,hu2020heterogeneous}, making it hard for those methods to generalize across different heterogeneous graphs. Moreover, learning node representations purely from pre-defined meta paths easily makes a model overlook the important first-order (i.e., one-hop) neighbor nodes, since meta paths are more concerned on nodes connected in two hops or further distances \cite{park2020meta}. For instance, when learning a user's preference for recommendation, all her/his interacted item nodes carry strong indicative signals on the user's personal interests. Unfortunately, such information has to be sacrificed in a meta path-based learning scheme, leading to considerable amount of information loss in the produced node embeddings. 

In this regard, another popular solution for heterogeneous graph representation learning is built upon the neighborhood message passing paradigm as in homogeneous GNNs, while infusing additional information about the heterogeneity of the graph. For example, given all sampled neighbors of a target node, both RGCN \cite{schlichtkrull2018modeling} and HGT \cite{hu2020heterogeneous} design a type-specific aggregation scheme to firstly aggregate the information within each type of neighbor nodes, then pass all summarized messages to the target node via a second-tier aggregator. Though promising performance is reported, such neighborhood-based aggregation scheme is highly vulnerable to sparse graphs where most nodes are loosely connected. Take the widely used user-item recommendation graphs (e.g., Amazon \cite{kang2018self} and Yelp \cite{nilizadeh2019think}) as an example, the average degree of each user node is commonly below 5, leading to a limited amount of messages available for learning user node embeddings. Though methods based on GraphSAGE \cite{hamilton2017inductive,wang2019neural,ying2018graph} can grasp extra information from high-order neighbor nodes by increasing the neighborhood sampling depth, it brings an inevitable by-product of recursively increasing computational cost and memory footprint \cite{chen2018fastgcn} to cope with the expanded neighborhood. 

Meanwhile, a large body of previous works \cite{kipf2017GCN,yun2019graph,velivckovic2018graph} on GNNs are dedicated to embedding nodes in a fixed graph, e.g., GCN-based approaches \cite{kipf2017GCN} require a full adjacency matrix of the graph. That is to say, they are based on a transductive learning paradigm, where all nodes must be present during training to facilitate downstream tasks. However, in real-world scenarios, many applications require embeddings to be quickly generated for unseen nodes. Such capability is essential for high-throughput, production machine learning systems \cite{hamilton2017inductive}, most of which involve heterogeneous graphs \cite{wang2019kgat,ying2018graph,gao2018bine}. The transductive learning scheme hurts the generalizability of those methods as they have to be retained once the graph is updated. Hence, the inductive learning capability is highly desirable for heterogeneous GNNs, which benefits representation learning on evolving graphs that constantly encounter unseen nodes, e.g., new users and videos on Youtube. Recently, attempts have been made to account for both inductiveness and heterogeneity in GNNs \cite{hu2020heterogeneous,wang2019heterogeneous}, but those methods heavily rely on sophisticated network structures (e.g., the meta path-specific attention network in HAN \cite{wang2019heterogeneous} and the hierarchical transformer in HGT \cite{hu2020heterogeneous}) in order to fully capture the contexts associated with different node/edge types. Therefore, practicality-wise, compared with the simplistic homogeneous GNNs, existing heterogeneous GNNs exhibit compromised training efficiency due to the complex architectures and an excessive amount of parameters to be learned. 

In light of the outstanding problems from heterogeneity, inductiveness and efficiency perspectives in graph representation learning, we aim to thoroughly investigate all these three issues in a unified view. Specifically, we propose \textbf{wi}de and \textbf{de}ep message passing \textbf{n}etwork (WIDEN) as an inductive and efficient solution to heterogeneous graph representation learning. To ensure inductiveness, WIDEN essentially follows the message passing paradigm in GNNs, where information is propagated from sampled neighbor nodes to the target node to form a node's embedding. To account for the heterogeneity of graphs, instead of using the inflexible meta paths or the cumbersome relation-specific aggregation schemes, we propose a novel heterogeneous message packaging paradigm that does not ask for seperate modelling of different types of meta paths/relations. Specifically, in a sampled neighborhood, we generate heterogeneous message packs by posing interactions between a neighbor node and its associated edge. As message packs are concretized representations of heterogeneous information passed from neighbor nodes, the forward propagation in WIDEN is performed at the message pack level and can thoroughly capture the heterogeneity of the graph. 

To overcome the aforementioned dilemma of utilizing only meta paths or local neighborhood for heterogeneous message passing, WIDEN specifies two types of neighborhoods, i.e., wide and deep neighborhoods. Essentially, the wide neighborhood and deep neighborhood can be respectively viewed as the width-first search node set and random walk node sequence.
%we devise a novel \textbf{wide and deep message passing} scheme in WIDEN. 
In a nutshell, when learning the embedding for a target node, the new message passing paradigm simultaneously makes full use of its local neighborhood information, and ensures sufficient outreach to further nodes. 
%With an inductive node embedding paradigm, WIDEN specifies two types of neighborhoods, i.e., wide and deep neighborhoods. Essentially, the wide neighborhood and deep neighborhood can be respectively viewed as the width-first search node set and random walk node sequence. By propagating the information from both wide and deep neighbor nodes to the target node, we can compensate for meta path-based methods' omission on low-order neighbors that tend to have higher relevance to the target node, while ensuring sufficient learning signals from the long-range node sequence when low-order neighbors are scarce. 
Besides, we introduce an innovative downsampling scheme in WIDEN to boost its training efficiency. Intuitively, for each target node, we prune its wide and deep neighbor nodes during training based on a learnable attentive weight indicating each neighbor's relevance to the target node. Notably, we further propose a specialized pruning method for deep neighbor sequences to avoid losing important semantic information if nodes are directly removed from a complete path.

To conclude, we make the following contributions:
\begin{itemize}
	\item We systematically investigate the heterogeneity, inductiveness, as well as efficiency when learning representations for graphs in a unified view. We propose WIDEN, which is an innovative graph neural network variant that supports inductive and efficient representation learning on heterogeneous graphs. 
	\item The novel message packaging paradigm, coupled with our wide and deep message passing architecture, makes WIDEN free of manually crafted meta paths and complex relation-specific modelling, while ensuring a balanced blend of messages from local neighborhood and high-order connections. The proposed downsampling strategy for both wide and deep neighbor nodes further facilitates efficient training of WIDEN with minimal accuracy loss.
	\item We conduct extensive experiments on three real-world heterogeneous graphs, and the results show that WIDEN outperforms state-of-the-art baselines in both transductive and inductive representation learning tasks. Furthermore, WIDEN is highly efficient in terms of time consumption during training.
\end{itemize}

\section{Preliminaries}\label{sec:pre}
In this section, we present the key definitions that are frequently used in our paper.

\textbf{\textit{Definition 1: Heterogeneous Graph.}} A heterogeneous graph $\mathcal{G} = \{\mathcal{V}, \mathcal{E}\}$ consists of a node set $\mathcal{V}$ and an edge set $\mathcal{E}$. Mathematically, an edge between nodes $v_i$, $v_j \in \mathcal{V}$ is defined as $e_{ij} = (v_i, v_j)$ where both nodes and edges are heterogeneous, e.g., authorship relation between author and paper nodes, or citation relation between two paper nodes in an academic graph. %where edge $e_{ij}$ also corresponds to an one-hot encoding of edge type

\textbf{\textit{Definition 2: Wide Neighbor Node Set.}} For a given target node $v_t\in \mathcal{V}$, its wide neighbor node set is defined as $\mathcal{W}(v_t) = \{(n,i)\}_{n=1}^{N_w}$ that stores the \textbf{indexes} $i$ of its uniformly sampled first-order neighbor nodes, and $|\mathcal{W}(v_t)| = N_w$ where $N_w$ is the initial sample size. In each tuple $(n,i)$, $i\in [1, |\mathcal{V}|]$ is the \textbf{global index} that is unique for each node $v_i \in \mathcal{V}$, while $n \in  [1, N_w]$ is the \textbf{local index} used only inside the set $\mathcal{W}(v_t)$.

\textbf{\textit{Definition 3: Deep Neighbor Node Set.}} For target node $v_t\in \mathcal{V}$, its deep neighbor set $\mathcal{D}(v_t)$ is initialized by performing a deep random walk of length $N_d$ that starts from $v_t$. Hence, $\mathcal{D}(v_t)=\{(s,i)\}_{s=1}^{N_d}$ where $s \leq N_d$ and $i \leq |\mathcal{V}| = N_d$ are respectively the local and global node indexes. Note that we do not include the target node itself in both neighbor sets. In $\mathcal{D}(v_t)$, $s$ is exactly the position of nodes in the sequence generated by the random walk, e.g., $s=1$ and $s=N_d$ correspond to the first node (i.e., one of $v_t$'s first-order neighbor) and last node in the sequence, respectively. 

\textbf{Notations.} Throughout this paper, all vectors are \textbf{row vectors} unless specified, e.g., $\textbf{v}_t \in \mathbb{R}^{1\times d}$. To maintain simplicity, we use superscripts $\circ$ and $\triangleright$ to distinguish notations for wide and deep neighbors, respectively. To make the subscripts easy to follow across computations, we resort to \textbf{local indexes} when denoting the neighbor nodes of $v_t$, i.e., $v_n = v_i$ for $(n,i)\in \mathcal{W}(v_t)$ and $v_s = v_i$ for $(s,i)\in \mathcal{D}(v_t)$.

\textbf{Embedding Initialization.} Each node is associated with a $d_0$-dimensional feature vector $\textbf{x}_t \in \mathbb{R}^{1\times d_0}$ that is used to initialize its embedding, i.e., $\textbf{v}_t = \textbf{x}_t\textbf{G}^{node}$ with linear projection weight $\textbf{G}^{node}\in \mathbb{R}^{d_0 \times d}$. Also, for edge $e_{ij}$ between any two nodes, we define its type-specific embedding as $\textbf{e}_{ij} = \mathrm{onehot}(e_{ij})\textbf{G}^{edge}$, where $\mathrm{onehot}(\cdot)$ returns the one-hot encoding of $e_{ij}$'s type, and $\textbf{G}^{edge} \in \mathbb{R}^{d_0' \times d}$ is the embedding matrix for a total of $d_0'$ edge types.

\section{Wide and Deep Message Passing Network}
In a wide and deep message passing step, taking the current representation $\textbf{v}_t$ for the target node as the input, our inductive message passing scheme aims to generate an updated representation $\textbf{v}_t'$ by propagating information from both the wide and deep neighbor node sets of $v_t$. To thoroughly account for the heterogeneity of the graph, apart from the diversified node features, we take the specific type of the edges between nodes into account. Hence, for wide and deep neighbor sets, we first define two corresponding heterogeneous message packaging functions ${\pack}^{\circ}(\cdot)$ and ${\pack}^{\triangleright}(\cdot)$ to have all node and edge information packed up in matrices $\textbf{M}^{\circ}_t$ and $\textbf{M}^{\triangleright}_t$. The generated message packs diverges from meta path-based methods that do not specifically model edge information. Then, we propose two information passing functions ${\pass}^{\circ}(\cdot)$ and ${\pass}^{\triangleright}(\cdot)$ that respectively aggregate the information within the wide and deep message packs into two compact vectors $\textbf{h}^{\circ}_t$ and $\textbf{h}^{\triangleright}_t$. Afterwards, function $\fuse(\cdot)$ fuses both representations and generates $\textbf{v}_t'$ to replace the original node embedding. 

Note that as both wide and deep neighbor sets are subject to downsampling in WIDEN, we use $|\mathcal{W}(v_t)|$ and $|\mathcal{D}(v_t)|$ to respectively denote the current number of neighbors in each set instead of using the fixed initial neighbor sizes $N_w$ and $N_d$. The downsampling phase essentially reduces the amount of messages need to be passed into $v_t$, thus allowing for more efficient training. In what follows, we present the design of those key components in detail.

%\begin{algorithm}
%\caption{A Wide and Deep Message Passing Step}\label{Algorithm:WDMP}
%\begin{algorithmic}[1]
%\State \textbf{Input:} Current representation $\textbf{v}_t$ for target node $v_t \in \mathcal{V}$, wide and deep message packaging functions $\pack^{\circ}(\cdot)$ and $\pack^{\triangleright}(\cdot)$, wide and deep message passing functions $\pass^{\circ}(\cdot)$ and $\pass^{\triangleright}(\cdot)$;
%\State \textbf{Output:} Updated vector representation $\textbf{v}_t'$ for $v_t$
%\State $\textbf{M}^{\circ}_t \leftarrow \pack^{\circ}(\mathcal{W}(v_t))$, $\textbf{M}^{\triangleright}_t \leftarrow \pack^{\triangleright}(\mathcal{D}(v_t))$;
%\State $\textbf{h}_t^{\circ} \leftarrow \pass^{\circ}(\textbf{M}^{\circ}_t)$, $\textbf{h}_t^{\triangleright} \leftarrow \pass^{\triangleright}(\textbf{M}^{\triangleright}_t)$;
%\State $\textbf{v}_t' \leftarrow \fuse (\textbf{h}_t^{\circ}, \textbf{h}_t^{\triangleright})$;
%\State \textbf{end}
%\end{algorithmic}
%\end{algorithm}

\begin{figure}[!t]
\vspace{-0.2cm}
\center
\includegraphics[width = 3.4in]{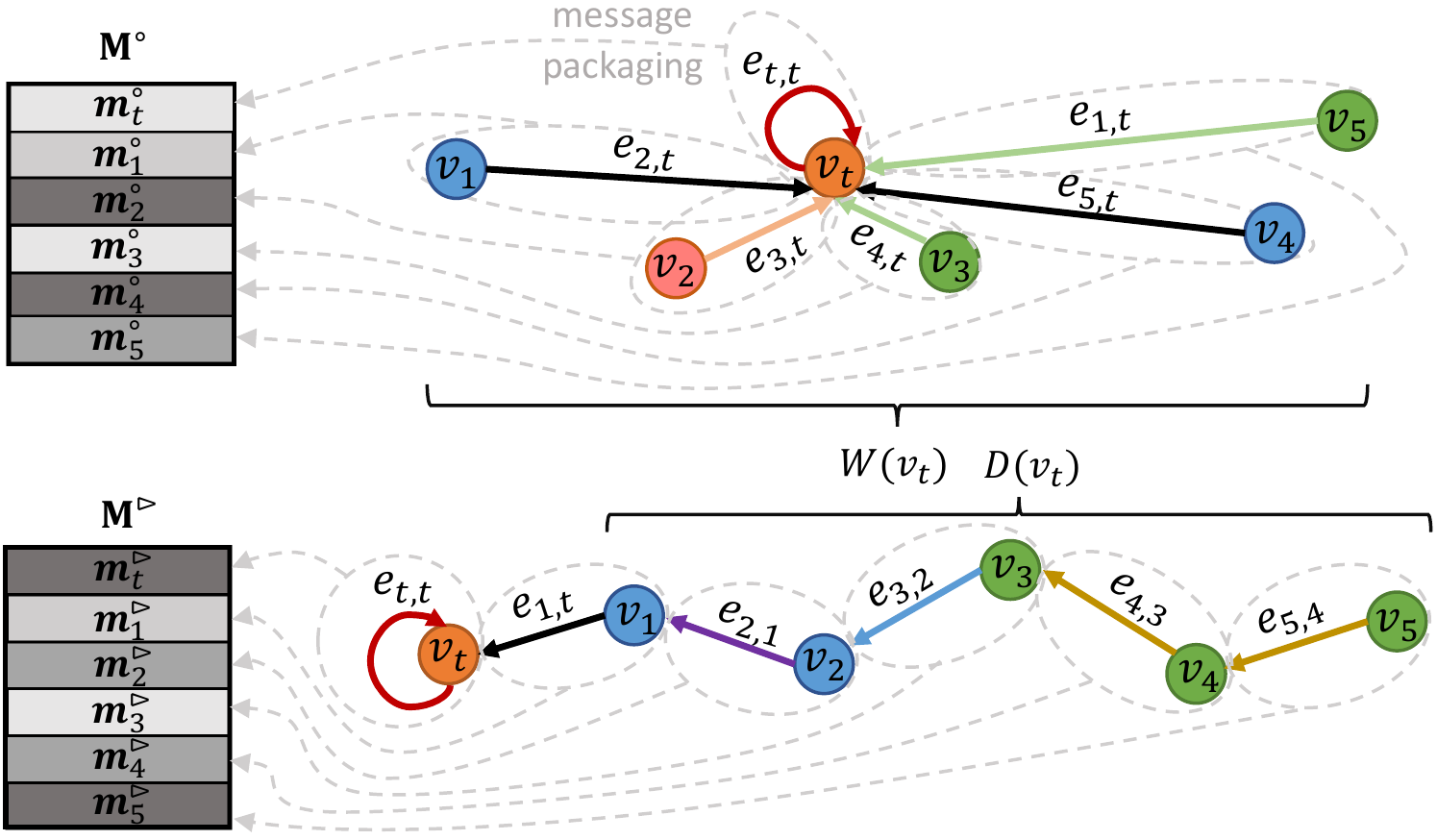}
\vspace{-0.4cm}
\caption{A schematic view of message packaging in WIDEN. Top and bottom parts correspond to wide and deep neighbor sets, respectively. In this figure, we suppose $|\mathcal{W}(v_t)| = |\mathcal{D}(v_t)| = 5$ for illustration purposes.}
\label{Figure:MP}
\vspace{-0.4cm}
\end{figure}

\subsection{Heterogeneous Message Packaging}
As defined in Section \ref{sec:pre}, the wide neighbor set $\mathcal{W}(t)$ contains $|\mathcal{W}(v_t)|$ sampled first-order neighbors for target node $v_t$. To fully preserve the heterogeneous information within $\mathcal{W}(t)$ during message passing, we define the following message packaging process:
\begin{align}\label{eq:wide_packaging}
	\textbf{M}^{\circ} & = {\pack}^{\circ}(\mathcal{W}(v_t)) \nonumber\\
	& =	
	\begin{bmatrix}
	\textbf{m}_t^{\circ} \\
    \textbf{m}_1^{\circ} \\
    \vdots \\
    \textbf{m}_{|\mathcal{W}(v_t)|}^{\circ}
    \end{bmatrix} = 
    \begin{bmatrix}
    \textbf{v}_t \odot \textbf{e}_{t,t}\\
 	\vertconcat_{n=1}^{|\mathcal{W}(v_t)|} \!\!\!\!\!\!\!\! \underbrace{(\textbf{v}_n \odot \textbf{e}_{n,t})}_{\!\!\!\!\!\!\!\!\!\!\! \textnormal{\small a single message pack}}\!
 	\end{bmatrix},
\end{align}
where $\odot$ is the element-wise multiplication, and $\parallel$ denotes the iterative operator for vertical concatenation of vectors. As illustrated in Figure \ref{Figure:MP}, the resulted matrix $\textbf{M}^{\circ} \in \mathbb{R}^{(|\mathcal{W}(v_t)| + 1)\times d}$ carries the message packs of the target node itself $\textbf{m}_t^{\circ}$ and all its wide neighbors $\textbf{m}_1^{\circ},...,\textbf{m}_{|\mathcal{W}(v_t)|}^{\circ}$. Specifically, $\textbf{e}_{n,t} \in \mathbb{R}^{1\times d}$ denotes the trainable embedding vector of the relation (i.e., edge type) between the target node $v_t$ and neighbor $v_n$. For example, when modelling academic graphs, the edge types may include scholar-paper authorship, paper-paper citation, paper-topic subordination, etc. After the element-wise interaction between the node representation and edge embedding, the resulted $d$-dimensional vector encodes the heterogeneous information of both parts, which we term a \textbf{\textit{message pack}} $\textbf{m}$. Notably, for $\textbf{m}^{\circ}_t$, i.e., the target node $v_t$, we also learn a self-loop edge embedding $\textbf{e}_{t,t}$ between the same type of nodes (e.g., conference-conference self-loop edge).

Similarly, as Figure \ref{Figure:MP} depicts, we can construct a deep message matrix $\textbf{M}^{\triangleright}\in \mathbb{R}^{(|\mathcal{D}(v_t)|+1)\times d}$ by stacking all heterogeneous message packs obtained from $\mathcal{D}(v_t)$:
\begin{align}\label{eq:deep_packaging}
	\textbf{M}^{\triangleright} &= {\pack}^{\triangleright}(\mathcal{D}(v_t)) \nonumber\\ 
	&= 	
	\begin{bmatrix}
    \textbf{m}_t^{\triangleright} \\
    \textbf{m}_1^{\triangleright} \\
    \vdots \\
    \textbf{m}_{|\mathcal{D}(v_t)|}^{\triangleright}
    \end{bmatrix} = 
        \begin{bmatrix}
    \textbf{v}_t \odot \textbf{e}_{t,t}\\
 	\vertconcat_{n=1}^{|\mathcal{D}(v_t)|}  (\textbf{v}_s \odot \textbf{e}_{s,s-1})
 	\end{bmatrix},
\end{align}
where except for $\textbf{m}_{t}^{\triangleright}$, each message pack is formed by fusing the $s$-th node representation $\textbf{v}_s$ with edge $\textbf{e}_{s,s-1}$ linking its predecessor $v_{s-1}$ in the random walk sequence. Note that $\textbf{e}_{1,0}=\textbf{e}_{1,t}$ connects the first node in $\mathcal{D}(v_t)$ with $v_t$.

\subsection{Inductive Message Passing}
We hereby introduce the message passing scheme in WIDEN that coordinates heterogeneity and inductivity for node representation learning. 

\textbf{Wide Attentive Message Passing.} Intuitively, given $\mathcal{W}(v_t)$, we learn the representation of target node $v_t$ by modelling the combinatorial message passed from its neighbors as well as itself. Among the literatures of homogeneous graphs, the most common message passing scheme is defined in \cite{hamilton2017inductive}, where an aggregator (e.g., average pooling) is applied for merging the information from $v_t$ itself and its neighborhood. However, as pointed out in \cite{velivckovic2018graph}, instead of treating all neighbor nodes equally, different neighbor nodes should be weighted proportionally based on their importance to target node $v_t$. In the context of heterogeneous graphs, the capability of distinguishing the varied contributions from all heterogeneous message packs becomes especially crucial owing to the diversity of node and edge semantics. Hence, in WIDEN, we devise the following self-attention unit \cite{vaswani2017attention} to selectively transfer the information within $\textbf{M}^{\circ}$ to $v_t$:
\begin{equation}\label{eq:wideMP}
\begin{split}
	\textbf{h}_t^{\circ} &= {\pass}^{\circ}(\textbf{M}^{\circ}, \textbf{m}_t^{\circ}) \\
	&= \mathrm{softmax} \Big{(}\frac{\textbf{m}_t^{\circ}\textbf{W}^{\circ}_Q \cdot (\textbf{M}^{\circ}\textbf{W}^{\circ}_K)^{\top}}{\sqrt{d}} \Big{)}\cdot \textbf{M}^{\circ}\textbf{W}^{\circ}_V,
\end{split}
\end{equation}
where $\textbf{W}^{\circ}_Q$, $\textbf{W}^{\circ}_K$, $\textbf{W}^{\circ}_V \in \mathbb{R}^{d\times d}$ are respectively the query, key, and value weight matrices, while $\sqrt{d}$ is the scaling factor to smooth the row-wise softmax output and avoid extremely large values of the inner product. Note that we only take the message pack of target node, i.e., $\textbf{m}_t^{\circ}$ as the query of Eq.(\ref{eq:wideMP}), hence the resulted $|\mathcal{W}(v_t)|+1$ attentive weights $\{a^{\circ}_{t,t},a^{\circ}_{t,1},a^{\circ}_{t,2},...,a^{\circ}_{t,|\mathcal{W}(v_t)|}\}=\mathrm{softmax} (\frac{\textbf{v}_t\textbf{W}^{\circ}_Q \cdot (\textbf{M}^{\circ}\textbf{W}^{\circ}_K)^{\top}}{\sqrt{d}})$ carries the probability distribution indicating the importance of each heterogeneous message pack in $\textbf{M}^{\circ}_t$ to $v_t$. Then, through a weighted sum of all message packs, the self-attention unit summarized message for $v_t$, denoted by $\textbf{h}_t^{\circ}$. %The message packs in $\textbf{M}^{\circ}$ already encodes heterogeneous semantics from both nodes and edges
%Secondly, for most meta-path models, the learned node embeddings are purely conditioned on the node features only, making the heterogeneous edge information largely neglecte
Unlike existing relation-specific or meta path-specific graph embedding paradigms \cite{hu2020heterogeneous,wang2019heterogeneous,yun2019graph},  our self-attentive scheme avoids the need for repeating the computationally expensive aggregations for each pre-defined relations/meta paths and merging all information afterwards. As the message packs in $\textbf{M}^{\circ}$ essentially encodes heterogeneous semantics from both nodes and edges, $\textbf{h}_t^{\circ}$ is already an expressive representation of the heterogeneous connectivity around $v_t$.

\textbf{Successive Self-Attention for Deep Message Passing.} Learning node embeddings via sampled node sequences (i.e., paths) involves two widely used strategies. One originates from SkipGram \cite{mikolov2013efficient}, where the core is to maximize the log-likelihood of observing a sampled homogeneous node path \cite{perozzi2014deepwalk,grover2016node2vec}. The other adopts recurrent neural networks (RNNs) which allows a node's message to be propagated to distantly connected nodes, and is proven effective in capturing the information cascade within node sequences \cite{li2017deepcas,yang2019multi}. Speaking of heterogeneity, the majority of state-of-the-art message passing networks are based on the notion of meta paths. However, meta path-based approaches are heavily constrained by their inflexibility and high dependent of domain expertise. To overcome this dilemma, we propose to subsume $\textbf{M}^{\triangleright}$, which is essentially a sequence of the deep message packs, under a sequential modelling paradigm with our proposed successive self-attentive message passing operation:
\begin{equation}\label{eq:deepMP1}
	\textbf{H}^{\triangleright} = \mathrm{softmax} \Big{(}\frac{\textbf{M}^{\triangleright}\textbf{W}^{\triangleright}_Q \cdot (\textbf{M}^{\triangleright}\textbf{W}^{\triangleright}_K)^{\top}}{\sqrt{d}} + \Theta \Big{)}\cdot \textbf{M}^{\triangleright}\textbf{W}^{\triangleright}_V,
\end{equation}
which is followed by:
\begin{equation}\label{eq:deepMP2}
\begin{split}
	\textbf{h}_t^{\triangleright} &= {\pass}^{\triangleright}(\textbf{H}^{\triangleright}, \textbf{m}^{\triangleright}_{t}) \\
	&= \mathrm{softmax} \Big{(}\frac{\textbf{m}_t^{\triangleright}\textbf{W}^{\triangleright'}_Q \cdot (\textbf{H}^{\triangleright}_t\textbf{W}^{\triangleright'}_K)^{\top}}{\sqrt{d}}\Big{)}\cdot \textbf{M}^{\triangleright}\textbf{W}^{\triangleright'}_V.
\end{split}
\end{equation}

Essentially, the first self-attention in Eq.(\ref{eq:deepMP1}) updates each deep message pack in $\textbf{M}^{\circ}$ by capturing the sequential dependencies when information is propagated from the end of the sequence (i.e., $v_{|\mathcal{D}(v_t)|}$) to $v_t$. This provides the sequential characteristics of RNNs for node sequence modelling \cite{li2017deepcas,chen2019air,chen2020sequence}, while offering a substantially lower computational complexity. Then, $\textbf{H}^{\circ}$ can be viewed as a refined version of deep message packs with awareness of sequentiality. With the joint effect of $\pass^{\triangleright}(\cdot, \cdot)$ in Eq.(\ref{eq:deepMP2}), the resulted representation $\textbf{h}^{\triangleright}_t \in \mathbb{R}^{1\times d}$ will in return gather richer semantic information from the deep neighbors into the target node. However, conventional self-attention allows each element in the sequence to receive information from both preceding and succeeding ones, which is an inappropriate assumption for message passing in graph modelling. To ensure the information propagation is one-directional in a sequence, we incorporate an attention mask $\Theta$ into Eq.(\ref{eq:deepMP1}). Each element $\theta_{row, col}\in\Theta$ is defined as:
\begin{equation}\label{eq:mask}
	\theta_{row, col} = \Bigg{\{}
	\begin{array}{c}
			\hspace{0.4cm}0, \hspace{0.2cm} \textnormal{if}\hspace{0.2cm} row \leq col \\
			\hspace{-0.6cm}-\infty, \hspace{0.2cm} \textnormal{otherwise} \\
		\end{array}.\\
\end{equation}

\textbf{Rationale of The Attention Mask}.
We denote the matrix product of the query and key matrices in Eq.(\ref{eq:deepMP1}) as $\textbf{A}$, i.e., $\textbf{A} \!\! = \!\! \frac{\textbf{M}^{\triangleright}\textbf{W}^{\triangleright}_Q \cdot (\textbf{M}^{\triangleright}\textbf{W}^{\triangleright}_K)^{\top}}{\sqrt{d}} \!\in\! \mathbb{R}^{(|\mathcal{D}(v_t)|+1) \times (|\mathcal{D}(v_t)|+1)}$. By adding the attention mask $\textbf{M}^{\triangleright}$, the interaction scores $a_{row,col} \in \textbf{A}$ %from $i+1$ 
turn to $-\infty$ if $row > col$ and the softmax-normalized attentive weights consequently approximate $0$. The interaction scores on other positions remain unchanged, ensuring that each message pack only yields influence to previous ones in the deep neighbor sequence.

\textbf{Fusing Wide and Deep Representations.} After obtaining $\textbf{h}^{\circ}_t$ and $\textbf{h}^{\triangleright}_t$, we combine the message passing results from both wide and deep neighbors to update node $v_t$'s embedding. It is worth noting that, to better compensate for the insufficiency of low-order neighbors, in practice we will sample $\Phi \geq 1$ deep neighbor sets for $v_t$. With the deep message passing scheme, one representation can be derived for each deep neighbor set, which is denoted by $\textbf{h}^{\triangleright}_{\phi}$ with $1\leq \phi \leq \Phi$. For all $\Phi$ learned message representations, we merge them into a unified representation with a simplistic average pooling operation that does not introduce excessive computational costs. Lastly, to generate the updated embedding for $v_t$, we send both wide and deep messages into a feed-forward network with a normalization operation:
\begin{align}\label{eq:fuse}
	\textbf{h}_t & = \textnormal{ReLU}(\textbf{W}[\textbf{h}_t^{\circ}; \frac{1}{\Phi}\sum_{\phi = 1}^{\Phi} \textbf{h}_{t(\phi)}^{\triangleright}] + \textbf{b}),\nonumber\\
	\textbf{v}_t' & = \frac{\textbf{h}_t}{||\textbf{h}_t||},
\end{align}
where $[\cdot;\cdot]$ denotes horizontal concatenation of two vectors, $\textbf{W}\in \mathbb{R}^{2d\times d}$ and $\textbf{b}\in \mathbb{R}^{1\times d}$ are respectively the weight and bias, and $\textbf{v}_t'$ is the output embedding of target node $v_t$ in this deep and wide message passing step.

\subsection{Efficient Training via Downsampling} 
To facilitate fast model learning, on top of the computationally lightweight self-attention network to facilitate efficient feed-forward, we propose a downsampling strategy to progressively shrink the number of message packs need to be propagated in every training iteration of WIDEN. Intuitively, by filtering out nodes (and connected edges) from $\mathcal{W}(v_t)$ and $\mathcal{D}(v_t)$ that make the least contribution to the learning of $\textbf{v}_t'$. In short, by reducing the amount of message to be passed to node $v_t$, we can speed up the training process.
%significantly reduce the major computational cost exerted by Eq.(\ref{eq:wideMP}), Eq.(\ref{eq:deepMP1}) and Eq.(\ref{eq:deepMP2}). 
Furthermore, with a selective downsampling process, the discarded message packs are of less importance in the current iteration.
% removes potential noise and emphasises the most informative message packs
As such, though less neighborhood information is involved for learning $\textbf{v}_t'$, downsampling  high-quality node embeddings can be guaranteed in WIDEN. 

\begin{algorithm}[b]
\caption{A Wide Message Shrinking Step}
\label{Algorithm:WideShrinking}
\begin{algorithmic}[1]
\State \textbf{Input:} Message matrix $\textbf{M}^{\circ}$ for node $v_t$, attentive weights $\{a^{\circ}_{t,t},a^{\circ}_{t,1},a^{\circ}_{t,2},...,a^{\circ}_{t,|\mathcal{W}(v_t)|}\}$ computed by Eq.(\ref{eq:wideMP});
\State \textbf{Output:} Updated message matrix $\textbf{M}^{\circ'}$;
\State $\textbf{a}^{\circ} \leftarrow \{a^{\circ}_{t,1},a^{\circ}_{t,2},...,a^{\circ}_{t,|\mathcal{W}(v_t)|}\}$; \hspace{1cm}  (exclude $v_t$ itself)
\State $n' \leftarrow \mathrm{argmin}(\textbf{a}^{\circ})$;
\State $\mathcal{W}'(v_t) = \emptyset$;
\State \textbf{for each} $(n, i) \in \mathcal{W}(v_t)$ \textbf{do}
\vspace{0.1cm}
\State \hspace{0.4cm} $\mathcal{W}'(v_t) \cup \bigg{\{}
	\begin{array}{c}
			\hspace{0.4cm}(n, i), \hspace{0.2cm} \textnormal{if}\hspace{0.2cm} n < n' \\
			\hspace{-0.25cm}(n-1, i), \hspace{0.2cm} \textnormal{if}\hspace{0.2cm} n > n' \\
		\end{array}$
\vspace{0.1cm}
\State \textbf{end for}\hspace{1.2cm}  (update $\mathcal{W}(v_t)$ with new local indexes)
\State $\textbf{M}_{t}^{\circ'} \leftarrow \pack^{\circ}(\mathcal{W}'(v_t))$;
\end{algorithmic}
\end{algorithm}

\textbf{Shrinking Wide Message Packs.} In WIDEN, we advocate to fully utilize the attentive weights that are assigned to each neighbor node in $\mathcal{W}(v_t)$ for downsampling. Specifically, a single step for shrinking wide messages $\textbf{M}^{\circ}$ is presented in Algorithm \ref{Algorithm:WideShrinking}. Recall that $\{a^{\circ}_{t,t},a^{\circ}_{t,1},a^{\circ}_{t,2},...,a^{\circ}_{t,|\mathcal{W}(v_t)|}\}=\mathrm{softmax} (\frac{\textbf{v}_t\textbf{W}^{\circ}_Q \cdot (\textbf{M}^{\circ}\textbf{W}^{\circ}_K)^{\top}}{\sqrt{d}})$ are the $|\mathcal{W}(v_t)|+1$ attentive weights indicating the importance of message packs $\textbf{m}^{\circ}_t,\textbf{m}^{\circ}_1,\textbf{m}^{\circ}_2,...,\textbf{m}^{\circ}_{|\mathcal{W}(v_t)|}$ towards the learning of context $\textbf{h}^{\circ}_t$. Hence, after a message passing step, taking $\textbf{M}^{\circ}$ and the attentive weights as the input, Algorithm \ref{Algorithm:WideShrinking} actively removes $v_t$'s one wide neighbor $v_n \in \mathcal{W}(v_t)$ that has the smallest attentive score $a^{\circ}_{t,n}$.

\begin{figure}[!t]
\vspace{-0.2cm}
\center
\includegraphics[width = 3.6in]{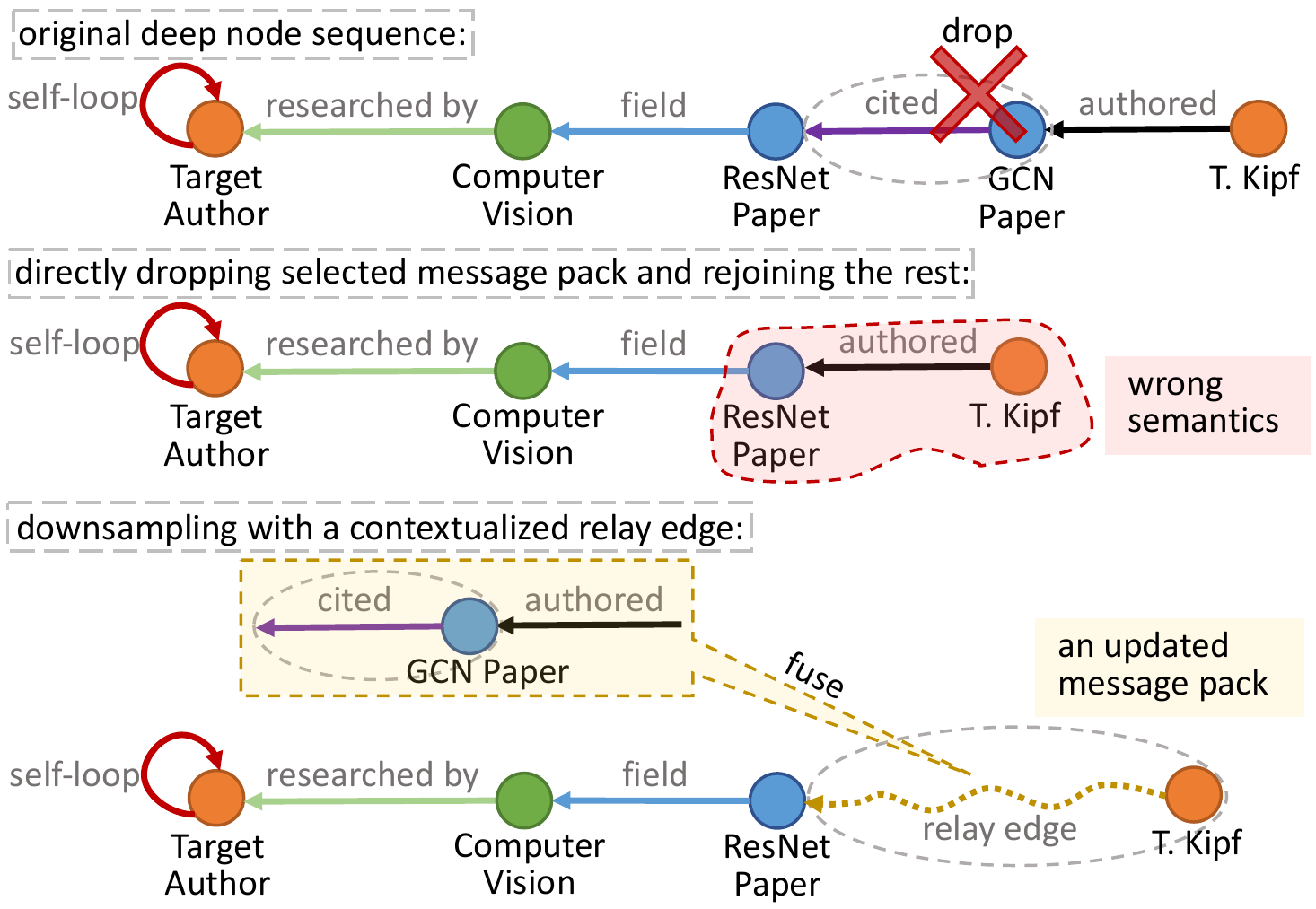}
\vspace{-0.8cm}
\caption{A demonstration of why directly dropping message packs from deep neighbors causes semantic loss, and how our contextualized relay edges alleviate this issue when downsampling. The ``Target Author'' node is $v_t$ in this figure.}
\label{Figure:relay}
\vspace{-0.4cm}
\end{figure}

\textbf{Pruning Deep Message Packs with Contextualized Relay Edges.}
For notation simplicity, we avoid individually referring to each $\phi$-th deep neighbor set of $v_t$ but keep using the single message matrix $\textbf{M}^{\triangleright}$ in this section. The basic idea behind deep message pack downsampling is similar to the attentive weight-based method for wide message packs. However, an important difference of message passing with $\mathcal{D}(v_t)$ against $\mathcal{W}(v_t)$ is that, the node neighbors form a long-range sequence that carries a complete, high-order context to be transmitted into $v_t$. Hence, straightforwardly deleting node and its associated edge (i.e., a message pack) will lead to the connectivity on the node sequence ill-posed as the modified node sequence no longer reflects how the message is passed from the last node $v_{|\mathcal{D}(v_t)|}$ to $v_t$ via heterogeneous relationships. As shown in Figure \ref{Figure:relay}, when learning representation for the target author node, directly dropping the message pack with ``GCN Paper'' node and ``cited'' edge and rejoining the remaining content in the node sequence will create a path ``T. Kipf authored ResNet Paper'', which carries wrong semantics and will impede the correctness of messages passed to $v_t$. 

\begin{algorithm}[t]
\caption{A Deep Message Pruning Step}
\label{Algorithm:DeepPruning}
\begin{algorithmic}[1]
\State \textbf{Input:} Message matrix $\textbf{M}^{\triangleright}$ for node $v_t$, attentive weights $\{a^{\triangleright}_{t,t},a^{\triangleright}_{t,1},a^{\triangleright}_{t,2},...,a^{\triangleright}_{t,|\mathcal{D}(v_t)|}\}$ computed by Eq.(\ref{eq:deepMP2});
\State \textbf{Output:} Updated message matrix $\textbf{M}^{\triangleright'}$;
\State $\textbf{a}^{\triangleright} \leftarrow \{a^{\triangleright}_{t,1},a^{\triangleright}_{t,2},...,a^{\triangleright}_{t,|\mathcal{D}(v_t)|}\}$; \hspace{1cm}  (exclude $v_t$ itself)
\State $s' \leftarrow \textnormal{argmin}(\textbf{a}^{\triangleright})$;
\State \textbf{if} $s' < |\mathcal{D}(v_t)|$ \textbf{do}
\State \hspace{0.4cm} $\textbf{m}_{s'+1}^{\triangleright} \!\leftarrow\!$ Eq.(\ref{eq:relay}) where $\textbf{m}_{s'+1}^{\triangleright} \in \textbf{M}^{\triangleright}$; 
\State \textbf{end if}\hspace{1.2cm}(renew the message pack with relay edge)
\State $\mathcal{D}'(v_t) = \emptyset$;
\State \textbf{for each} $(s, i) \in \mathcal{D}(v_t)$ \textbf{do}
\vspace{0.1cm}
\State \hspace{0.4cm} $\mathcal{D}'(v_t) \cup \bigg{\{}
	\begin{array}{c}
			\hspace{0.4cm}(s, i), \hspace{0.2cm} \textnormal{if}\hspace{0.2cm} s < s' \\
			\hspace{-0.25cm}(s-1, i), \hspace{0.2cm} \textnormal{if}\hspace{0.2cm} s > s' \\
		\end{array}$
\vspace{0.1cm}
\State \textbf{end for}\hspace{1.2cm}  (update $\mathcal{D}(v_t)$ with new local indexes)
\State $\textbf{M}_{t}^{\triangleright'} \leftarrow \pack^{\triangleright}(\mathcal{D}'(v_t))$;
\State $\textbf{m}_{s'}^{\triangleright'} \leftarrow \textbf{m}_{s'+1}^{\triangleright}$ where $\textbf{m}_{s'}^{\triangleright'} \in \textbf{M}^{\triangleright'}$;
\Statex (update $s'$-th message pack with relay edge information)
\end{algorithmic}
\end{algorithm}

In WIDEN, we propose to generate contextualized relay edges in order to fully preserve the semantic information within $\mathcal{D}(v_t)$ during the downsampling process. 
To be specific, let $\textbf{m}^{\triangleright}_{s'}$ be the message pack we will delete from $\textbf{M}^{\triangleright}$ ($s' \in \{1, 2, ..., |\mathcal{D}(v_t)|\}$). If $\textbf{m}^{\triangleright}_{s'}$ is not the last element in the deep node sequence (i.e., $s'<|\mathcal{D}(v_t)|$), then we need to compute an additional relay edge to preserve the context of $\textbf{m}_s$ before deleting it. As depicted in Figure \ref{Figure:relay}, to ensure that semantic information is retained for the entire path of $\mathcal{D}(v_t)$, we update its subsequent message pack $\textbf{m}_{s'+1}$ via the following:
\begin{equation}\label{eq:relay}
	\textbf{m}_{s'+1} \leftarrow \textbf{v}_{s'+1} \odot \underbrace{\mathrm{maxpool}(\textbf{e}_{s'+1,s'}, \textbf{m}_{s'})}_\textnormal{\small relay edge},
\end{equation}
where $\mathrm{maxpool}(\cdot,\cdot)$ is the element-wise maxpooling operation between two vectors. With Eq.(\ref{eq:relay}), we present Algorithm \ref{Algorithm:DeepPruning} for pruning one deep message pack in a single iteration. In short, the relay edge binds the useful information from both the deprecated message pack $\textbf{m}_s$ and the edge embedding $\textbf{e}_{s+1}$, thus generating a contextualized relay edge between node $v_{s+1}$ and $v_{s-1}$ prior to the deletion of $v_s$. Then, the new message pack $\textbf{m}_{s+1}$ can be computed by replacing the original $\textbf{e}_{s+1}$ with the relay edge. Intuitively, the relay edge greatly helps minimize the loss of semantics while still enabling deep message pack downsampling to facilitate efficient training of WIDEN.

\subsection{Training WIDEN}\label{sec:opt}
We introduce the training process of WIDEN in this section.

\textbf{Training Process.}
Algorithm \ref{Algorithm:Training} describes the efficient training procedure of our model. In Algorithm \ref{Algorithm:Training}, lines 5-16 essentially iterate the mini-batch training by firstly engaging wide and deep message passing and then performing active downsampling for both neighbors. Noticeably, the downsampling process is subject to two additional constraints, i.e., the downsampling lower bounds $k^{\circ}$ and $k^{\triangleright}$ and the downsampling thresholds $r^{\circ}$ and $r^{\triangleright}$. The lower bounds $k^{\circ},k^{\triangleright}\geq 1$ controls the minimum neighbor set size that $v_t$ preserves during the downsampling process. Once there are only $k^{\circ}$/$k^{\triangleright}$ remain in the wide/deep neighbor set, WIDEN stops the downsampling so that neighbor nodes will no longer be dropped in future iterations. Meanwhile, the downsampling thresholds $r^{\circ}$ and $r^{\triangleright}$ determines circumstances where downsampling is triggered. We further elaborate on the downsampling triggering mechanism below.

\begin{algorithm}[t]
\caption{Training WIDEN}
\label{Algorithm:Training}
\begin{algorithmic}[1]
\State \textbf{Input:} Heterogeneous graph $\mathcal{G}=(\mathcal{V}, \mathcal{E})$, raw features of all nodes $\{\textbf{x}_t\}_{t=1}^{|\mathcal{V}|}$, neighborhood functions $\mathcal{W}(\cdot)$ and $\mathcal{D}(\cdot)$, initial neighbor sizes $N_w$ and $N_d$, batch size $B$, learning rate $\tau$, maximum training epoch $Z$, downsampling thresholds $r^{\circ}$ and $r^{\triangleright}$; downsampling lower bounds $k^{\circ}$ and $k^{\triangleright}$;
\State \textbf{Output:} Vector representations $\textbf{v}_t$ for all $v_t \in \mathcal{V}$;
\State Sample $\mathcal{W}(v_t)$ and $\mathcal{D}(v_t)$ for all $v_t \in \mathcal{V}$;
\State $\textbf{M}^{\circ} \leftarrow$ Eq.(\ref{eq:wide_packaging}), $\textbf{M}^{\triangleright} \leftarrow$ Eq.(\ref{eq:deep_packaging}), $z \leftarrow 0$; \hspace{1cm}(initialization)
\State \textbf{repeat}
\State \hspace{0.4cm} Draw $B$ target nodes from $\mathcal{V}$;
\State \hspace{0.4cm} \textbf{for each} target node $v_t$ in the training batch \textbf {do}  
\State \hspace{0.8cm} $\textbf{v}_t \leftarrow$ Eq.(\ref{eq:wideMP})-(\ref{eq:fuse});
\State \hspace{0.8cm} \textbf{if}  $z >1$ and $KL_{z}^{\circ} < r^{\circ}$ and $|\mathcal{W}(v_t)|>k^{\circ}$ \textbf{do}
\State \hspace{1.2cm} $\textbf{M}^{\circ} \leftarrow$ Algorithm \ref{Algorithm:WideShrinking};
\State \hspace{0.8cm} \textbf{end if}
\State \hspace{0.8cm} \textbf{if} $ z >1$ and $KL^{\triangleright}_{z} < r^{\triangleright}$ and $|\mathcal{D}(v_t)|>k^{\triangleright}$ \textbf{do}
\State \hspace{1.2cm} $\textbf{M}^{\triangleright} \leftarrow$ Algorithm \ref{Algorithm:DeepPruning};
\State \hspace{0.8cm} \textbf{end if} \hspace{1.3cm}(we assume $\Phi = 1$ to be succinct)
\State \hspace{0.4cm} $L \leftarrow$ Eq.(\ref{eq:loss});
\State \hspace{0.4cm} Take a gradient step to optimize $L$ with rate $\tau$;
\State \hspace{0.4cm} \textbf{if} the end of $\mathcal{V}$ is reached \textbf{do}
\State \hspace{0.8cm} $z \leftarrow z+1$;
\State \hspace{0.4cm} \textbf{end if}
\State \textbf{until} $L$ converges or $z=Z$;
\end{algorithmic}
\end{algorithm}

\textbf{Triggering Downsampling with KL Divergence.}
The downsampling process should be used with caution as suboptimal performance might be caused by the aggressively deleted data. This is because that if we do not allow sufficient training iterations to let the model ``warm up'' and grasp sufficient knowledge about a target node's neighbor sets, directly reducing the amount of node neighbors will likely lead to high instability in the training phase. A straightforward solution is to put the downsampling process on hold until WIDEN's training loss stabilizes, however this is a costly option due to the potentially large time consumption before the loss converges. Hence, we propose to dynamically quantify WIDEN's confidence in learning from $v_t$'s neighbors in the current training epoch. Taking the wide neighbor set as an example, for the same target node $v_t$, if the same $\mathcal{W}(v_t)$ is used for two consecutive epochs $z-1$ and $z$, then we can compare WIDEN's information gain in the current (i.e., $z$-th) over the previous (i.e., ($z-1$)-th) epoch from neighbor set $\mathcal{W}(v_t)$. Intuitively, as each epoch learns a probability distribution $\{a^{\circ}_{t,t},a^{\circ}_{t,1},a^{\circ}_{t,2},...,a^{\circ}_{t,|\mathcal{W}(v_t)|}\}$ over all $|\mathcal{W}(v_t)| + 1$ message packs, the information gain can be concretized by comparing the distributions learned in two epochs. As such, we formulate this process as computing the Kullback-Leibler (KL) divergence between two sets of attentive weights:
\begin{equation}\label{eq:KL}
	\!KL_{z}^{\circ} \!=\! 
	\begin{cases}
			a^{\circ}_{t,t({z-1})}\ln \frac{a^{\circ}_{t,t(z)}}{a^{\circ}_{t,t(z-1)}} + \sum_{n=1}^{|\mathcal{W}(v_t)|}a^{\circ}_{t,n(z-1)}\ln \frac{a^{\circ}_{t,n(z)}}{a^{\circ}_{t,n(z-1)}},\\
			\hspace{4cm} \textnormal{if}\hspace{0.2cm} \mathcal{W}_{z}(v_t) = \mathcal{W}_{z-1}(v_t) \\
			+\infty, \hspace{3.2cm}\textnormal{otherwise} \\
	\end{cases}
\end{equation}
where we append subscript $z$ or $z-1$ to differentiate notations for the current or previous training epoch. Notably, by feeding Eq.(\ref{eq:KL}) with $\{a^{\triangleright}_{t,t},a^{\triangleright}_{t,1},a^{\triangleright}_{t,2},...,a^{\triangleright}_{t,|\mathcal{D}(v_t)|}\}$, we can calculate the KL divergence regarding each deep neighbor set $KL^{\triangleright}_z$ at epoch $z$. In general, a sufficiently small $KL_z$ means that WIDEN acquires low information gain by performing attentive message passing on the same neighbor set again, and is hence highly confident in capturing information within the current neighbor set. Then, given $KL^{\circ}_z<r^{\circ}_z$ or $KL^{\triangleright}_z<r^{\triangleright}_z$, we can safely downsample the neighbors and let WIDEN re-capture the attentive weights with updated neighbors.

\textbf{Loss Function.}
As a versatile and generic heterogeneous graph embedding model, WIDEN can be optimized for different downstream tasks. In this paper, we focus on semi-supervised node classification, which is one of the most popular applications for heterogeneous graph embedding models \cite{yun2019graph,wang2019heterogeneous,hu2020heterogeneous}. Given the embedding $\textbf{v}_t'$ of an arbitrary node (e.g., an author in a citation network), we aim to predict its categorical label (e.g., the research field of an author). To optimize WIDEN towards the task goal, we enforce the model to output a probability distribution over all classes, and adopt cross-entropy to quantify the training error for Algorithm \ref{Algorithm:Training}:
\begin{equation}\label{eq:loss}
	 L = - \sum_{t\in \mathcal{Y}_{train}}{\textbf{y}_t^{\top} \log(\mathrm{softmax}(\textbf{v}_t' \textbf{C}))},
\end{equation}
where $\textbf{y}_t = \{0,1\}^{1\times c}$ is node node $v_t$'s one-hot label among $c$ different classes, $\mathcal{Y}_{train}$ is the full training set, while we use projection weight $\textbf{C} \in \mathbb{R}^{d\times c}$ followed by $\mathrm{softmax}(\cdot)$ to estimate $v_t$'s probability distribution over all classes from $\textbf{v}_t'$. In semi-supervised circumstances where we only have labels for a fraction of nodes, WIDEN iterates over all nodes in $\mathcal{V}$ while masking out unlabelled nodes in the graph.

\section{Experiments}\label{sec:exp}
In this section, we conduct experiments to evaluate WIDEN in terms of both effectiveness and efficiency. In particular, we aim to answer the following research questions (RQs) via experiments:
\begin{description}
	\item[\textbf{RQ1:}] How effectively can WIDEN perform heterogeneous graph representation learning?
	\item[\textbf{RQ2:}] Can WIDEN still learn expressive node representations in an inductive manner?
	\item[\textbf{RQ3:}] Is WIDEN efficient and scalable for training?
	\item[\textbf{RQ4:}] How WIDEN benefits from each component of the proposed model structure?
	\item[\textbf{RQ5:}] What is the impact of key hyperparameters to the performance of WIDEN? 
\end{description}

\subsection{Datasets}
We perform node classification task with three heterogeneous graph datasets, namely DBLP, ACM, and Yelp. The key statistics of our experimental datasets are listed in Table~\ref{table:Dataset}. DBLP and ACM are both academic graphs, and we use the source data from \cite{yun2019graph,wang2019heterogeneous}. Yelp\footnote{https://www.yelp.com/dataset} is a social business review graph that allows users to comment and rate different businesses. We briefly describe their characteristics below:
\begin{itemize}
	\item \textbf{DBLP}: In DBLP, the node types include \textit{paper}, \textit{author}, \textit{conference}, and \textit{terms}; and edge types include \textit{paper-author}, \textit{paper-conference}, and \textit{paper-term}. Each author node in DBLP is labeled by one of four research areas (i.e., \textit{database}, \textit{data mining}, \textit{machine learning}, and \textit{information retrieval}). Following \cite{yun2019graph,wang2019heterogeneous}, we use bag-of-words representations of each node's keywords as its raw features. 
	\item \textbf{ACM}: ACM contains \textit{paper}, \textit{author}, and \textit{subject} nodes, as well as \textit{paper-author} and \textit{paper-subject} edges. Labels are assigned to paper nodes based on the type of conferences (i.e., \textit{database}, \textit{wireless communication}, and \textit{data mining}) they are accepted to. Similar to DBLP, node features are formulated as bag-of-words representations.
	\item \textbf{Yelp}: Yelp is a million-scale heterogeneous graph with \textit{user}, \textit{business}, \textit{category} and \textit{attribute} nodes, where the edge types include \textit{user-business}, \textit{user-user}, \textit{business-category}, and \textit{business-attribute}. In Yelp, we label each business node's service quality as \textit{low}, \textit{medium}, and \textit{high}, which respectively corresponds to an overall rating below 3, between 3 and 4, and above 4. Such prediction is especially useful for evaluating new businesses where customer feedback is sparse. For user and business nodes, we extract Google's pre-trained word embeddings\footnote{https://code.google.com/archive/p/word2vec/} of all words from their associated reviews, which are then averaged as the input features. Word embeddings of keywords are also used as the raw features of category and attribute nodes. 
\end{itemize}

\begin{table}[!t]
\small
\caption{Statistics of datasets in use.}
\renewcommand{\arraystretch}{1.0}
\setlength\tabcolsep{3pt}
\center
  \begin{tabular}{c c c c c c}
    \toprule
    & \multirow{2}{*}{Property} & \multicolumn{3}{c}{Dataset} &\\
    \cline{3-5}
    & & ACM & DBLP & Yelp\\
	\midrule
	& \#Nodes & 8,994 & 18,405 & 2,179,470\\
	& \#Node Types & 3 & 4 & 4\\
	& \#Edges & 25,922 & 67,946 & 37,776,380\\
	& \#Edge Types & 2 & 3 & 4\\
	& \#Features & 1,902 & 334 & 184\\
	& \#Class Labels & 3 & 4 & 3\\
	\hline
	\multirow{3}{*}{\makecell{Transductive\\Learning}} & \#Training Nodes & 600 & 800 & 60,000\\
	& \#Validation Nodes & 300 & 400 & 30,000\\
	& \#Test Nodes & 2,125 & 2,857 & 119,392\\
	\hline
	\multirow{3}{*}{\makecell{Inductive\\Learning}} & \#Training Nodes & 2,420 & 3,245 & 167,513\\
	& \#Validation Nodes & - & - & -\\
	& \#Test Nodes & 605 & 812 & 41,879\\  
    \bottomrule
\end{tabular}
\label{table:Dataset}
\vspace{-0.6cm}
\end{table}

\subsection{Baseline Methods}
We compare WIDEN with the following state-of-the-art baseline methods:
\begin{itemize}
	\item \textbf{Node2Vec}: This method \cite{grover2016node2vec} extends the Word2Vec to the graph embedding task, which maximizes the co-occurrence probability among nodes from sequences generated from random walks.
	\item \textbf{GCN}: The graph convolutional network \cite{kipf2017GCN} introduces spectral convolutions on graphs to support semi-supervised representation learning.
	\item \textbf{FastGCN}: As a parallelizable model, FastGCN \cite{chen2018fastgcn} adopts a sampling strategy for mini-batch training while retaining similar performance as GCN.
	\item \textbf{GraphSAGE}: Using the sampling-and-aggregating scheme, GraphSAGE \cite{hamilton2017inductive} is a widely used message passing network for graph representation learning.
	\item \textbf{GAT}: The graph attention network  \cite{velivckovic2018graph} employs attention mechanism to selectively aggregate information from neighbor nodes based on their importance.
	\item \textbf{GTN}: The graph transformer network \cite{yun2019graph} first constructs meta-paths by multiplying relation-specific adjacency matrices, then performs graph convolutions on those transformed adjacency matrices. 
	\item \textbf{HAN}: By attentively selecting relevant information from the sampled meta-paths, the heterogeneous attention network \cite{wang2019heterogeneous} is currently the state-of-the-art heterogeneous graph embedding approach.
	\item \textbf{HGT}: With a relation-aware transformer model and a subgraph sampling strategy, the heterogeneous graph transformer \cite{hu2020heterogeneous} is the latest graph embedding model for balancing performance and efficiency.
\end{itemize} 

\begin{table*}[!t]
\small
\caption{Transductive node classification results. Different percentages indicate the proportion of labelled nodes used for training. Numbers in bold face are the best results in each column. Note that on Yelp, GTN incurs large training time cost (one epoch needs more than 10 hours) hence its performance is not reported. The results show that WIDEN significantly outperforms the best baseline in each column with $p$-value $< 0.05$ (underscored) or $p$-value $< 0.01$ (double-underscored) in paired $t$-test.}
\vspace{-0.3cm}
\centering
\renewcommand{\arraystretch}{1.0}
\setlength\tabcolsep{5pt}
  \begin{tabular}{|c|c|c|c|c||c|c|c|c||c|c|c|c|}
    \hline
    \multirow{2}{*}{Method} & \multicolumn{4}{c||}{ACM} & \multicolumn{4}{c||}{DBLP}  & \multicolumn{4}{c|}{Yelp}\\
    \cline{2-13}
    & 25\% & 50\% & 75\% & 100\% & 25\% & 50\% & 75\% & 100\% & 25\% & 50\% & 75\% & 100\%\\
    \hline
    Node2Vec \cite{grover2016node2vec} & 0.7797 & 0.7665 & 0.7906 & 0.7910 & 0.9077 & 0.9020 & 0.9140 & 0.9181 & 0.4098 & 0.4074 & 0.4060 & 0.4069 \\
    GCN \cite{kipf2017GCN} & 0.8058 & 0.8115 & 0.8133 & 0.8219 & 0.8381 & 0.8675 & 0.8683 & 0.8685 & 0.4523 & 0.4595 & 0.4765 & 0.4953 \\
    FastGCN \cite{chen2018fastgcn} & 0.7807 & 0.8875 & 0.9129 & 0.9188 & 0.7049 & 0.7871 & 0.8067 & 0.8039 & 0.5030 & 0.5265 & 0.6285 & 0.6638 \\
    GraphSAGE \cite{hamilton2017inductive} & 0.7567 & 0.7712 & 0.7816 & 0.8193 & 0.9062 & 0.9212 & 0.9205 & 0.9296 & 0.4629 & 0.5224 & 0.5475 & 0.5766 \\
    GAT \cite{velivckovic2018graph} & 0.8811 & 0.9076 & 0.9162 & 0.9128 & 0.8148 & 0.8443 & 0.8519 & 0.8667 & 0.4729 & 0.4748 & 0.5181 & 0.5208 \\
    GTN \cite{yun2019graph} & 0.8844 & 0.8835 & 0.8912 & 0.9021 & 0.9032 & 0.9044 & 0.9195 & 0.9310 & - & - & - & - \\
    HAN \cite{wang2019heterogeneous} & 0.8859 & 0.8950 & 0.9016 & 0.9052 & 0.9030 & \textbf{0.9219} & 0.9255 & 0.9247 & 0.4778 & 0.4910 & 0.5016 & 0.4832 \\
    HGT \cite{hu2020heterogeneous} & 0.8757 & 0.8875 & 0.9054 & 0.9089 & 0.7315 & 0.7693 & 0.7767 & 0.7778 & 0.4629 & 0.5224 & 0.5742 & 0.5940 \\
    \hline
    \textbf{WIDEN} & \underline{\textbf{0.8870}} & \underline{\textbf{0.9083}} & \underline{\underline{\textbf{0.9196}}} & \underline{\underline{\textbf{0.9269}}} & \underline{\textbf{0.9076}} & 0.9110 & \underline{\underline{\textbf{0.9293}}} & \underline{\underline{\textbf{0.9330}}} & \underline{\underline{\textbf{0.6892}}} & \underline{\underline{\textbf{0.7037}}} & \underline{\underline{\textbf{0.7125}}} & \underline{\underline{\textbf{0.7179}}}\\
     \addlinespace[0.03cm]
     \hline
    \end{tabular}
\label{table:transductive}
\vspace{-0.4cm}
\end{table*}

\subsection{Evaluation Protocols}\label{sec:eval_proto}
\textbf{Transductive Graph Representation Learning.} To avoid imbalance among node classes, we use \textbf{micro-averaged F1 score} to evaluate all methods' classification performance. We further vary the amount of training labels (i.e., 25\%, 50\%, 75\% and 100\% of the full training set) to test the robustness of each method with different supervision strengths. The split of training, validation, and test sets are given in Table~\ref{table:Dataset}.

\textbf{Inductive Graph Representation Learning.} We further evaluate the performance of all models that support inductive learning, i.e., a part of the full graph (containing labeled test nodes) is completely unseen during training. This setting is highly relevant to high-throughput production systems \cite{hamilton2017inductive} which constantly encounter unseen data and evolving graph structures. Hence, similar to \cite{hamilton2017inductive,velivckovic2018graph}, we randomly pick 20\%  labeled nodes from DBLP, ACM and Yelp, and exclude them from the graph during training. All model hyperparameters are inherited from the optimal configurations tuned with transductive setting, hence we no longer sample validation nodes in this experiment.

\subsection{Experimental Settings}\label{sec:settings}
To be consistent, we report the performance of WIDEN with a unified hyperparameter set $\{d=128, N_w = 20, N_d = 20, N_t = 10\}$. We also set the learning rate $\tau = 0.0001$, downsampling thresholds $r^{\circ}=r^{\triangleright}=0.001$ and lower bounds $k^{\circ}=k^{\triangleright}=5$. For DBLP and ACM, we adopt $L2$ regularization with strength $\gamma =0.01$ to prevent overfitting while we skip the regularization term on Yelp as the dataset is sufficiently large. The effect of different hyperparameter settings will be further discussed in Section \ref{sec:hypersens}. For all baselines, we optimize their hyperparameters via grid search with the validation set. Except for Node2Vec that is trained in a solely unsupervised manner, we adopt cross-entropy loss on the labeled training nodes to facilitate semi-supervised learning for all baselines. All effectiveness results are averaged over 5 executions. It is worth mentioning that some graph embedding baselines require the presence of the entire graph during training (i.e., Node2Vec, GCN, GAT, GTN and HAN), however the Yelp graph is too large to fit into the GPU memory. In this case, we adopt a well-established graph partitioning toolkit Metis \cite{karypis1998fast} to divide the full graph into multiple subgraphs by minimizing edge cuts, such that those models can iterate over all subgraphs in the training phase. 

\subsection{Effectiveness on Transductive Heterogeneous Graph Embedding (RQ1)} 
Table~\ref{table:transductive} reports the transductive node classification results of all methods on three datasets. It is worth noting that, GTN incurs huge time consumption when training on the largest dataset Yelp (one training epoch consumes more than $10$ hours) due to its CPU-only implementation, hence its performance is not reported on Yelp. Based on the transductive learning effectiveness, we draw the following observations.

First, in almost all cases, WIDEN consistently and significantly outperforms all the baseline methods. On the relatively small graphs ACM and DBLP, when $100\%$ of the labelled nodes in the training sets are used, WIDEN achieves a relative improvement of $1.98\%$ and $20.12\%$ over HGT, which is the latest heterogeneous GNN method. The margin becomes more apparent when tested on the large-scale graph Yelp with the full training set, where WIDEN advances the performance of the best homogeneous GNN (i.e., FastGCN) and heterogeneous GNN (i.e., HGT) by $8.15\%$ and $20.86\%$, respectively. This verifies that, by incorporating heterogeneous message passing schemes with both wide and deep neighborhoods, WIDEN is able to learn more expressive node representations for downstream tasks.

Second, homogeneous graph embedding methods does not always underperform on heterogeneous. For example, GAT and FastGCN respectively yield comparative performance on ACM and Yelp. A highly possible reason is the varied information richness regarding nodes and edges in a heterogeneous graph. Commonly the nodes are associated with sufficient raw features (e.g., user demographics) for learning high-quality representations while the edges connecting them do not. Consequently, compared with homogeneous methods, heterogeneous GNNs that additionally consider node and edge types sometimes show only marginal performance improvement. In contrast, by constructing message packs with the node and edge representations, WIDEN is capable of fully utilizing the heterogeneity of different edges for message passing.

Third, as we reduce the amount of labelled nodes for training, WIDEN demonstrates the slightest performance drop among all the methods. This is a highly desirable property in applications based on semi-supervised learning, as real-world data usually faces difficulties in acquiring sufficient training labels.

\begin{table}[t]
\small
\caption{Inductive node classification results. Numbers in bold face are the best results in each column. The results show that WIDEN significantly outperforms the best baseline on all datasets $p$-value $< 0.01$ (double-underscored) in paired $t$-test.}
\vspace{-0.3cm}
\centering
\renewcommand{\arraystretch}{1.0}
\setlength\tabcolsep{5pt}
  \begin{tabular}{|c|c|c|c|}
    \hline
    Method & ACM & DBLP & Yelp\\
       \hline
    GCN \cite{kipf2017GCN} & 0.5735 & 0.4921 & 0.3523 \\
    FastGCN \cite{chen2018fastgcn} & 0.5826 & 0.5237 & 0.3616 \\
    GraphSAGE \cite{hamilton2017inductive} & 0.8016 & 0.9185 & 0.4214 \\
    GAT \cite{velivckovic2018graph} & 0.9044 & 0.8543 & 0.5829 \\
    GTN \cite{yun2019graph} & 0.7829 & 0.8384 & - \\
    HAN \cite{wang2019heterogeneous} & 0.9005 & 0.9210 & 0.5315 \\
    HGT \cite{hu2020heterogeneous} & 0.9091 & 0.8264 & 0.6424 \\
    \hline
    \textbf{WIDEN}& \underline{\underline{\textbf{0.9175}}} & \underline{\underline{\textbf{0.9251}}} & \underline{\underline{\textbf{0.7613}}}\\
    \addlinespace[0.03cm]
     \hline
    \end{tabular}
\label{table:performance_inductive}
\vspace{-0.4cm}
\end{table}

\subsection{Effectiveness on Inductive Heterogeneous Graph Embedding (RQ2)}
We follow the evaluation protocols in Section \ref{sec:eval_proto} to test all methods' performance on inductive learning. Note that Node2Vec is excluded from this test as its design requires all node IDs to be known beforehand. For inductive GNNs (i.e., GCN, FastGCN and GTN), we mask out the corresponding nodes in the input feature matrix and adjacency matrix during training, and recover them for evaluation to approximate the effect of inductive learning. As inductivity has been an increasingly important aspect for graph-based applications, we evaluate WIDEN's performance in inductive settings both quantitatively and qualitatively. 

\textbf{Quantitative Results.} We list the quantitative classification results on inductively learned node embeddings in Table \ref{table:performance_inductive}. First of all, the inductive learning results place WIDEN in the leading position compared with all baselines. Hence, with the network parameters fully trained, WIDEN can effectively generalize to cold-start nodes that never appear in the training phase. Noticeably, as the inductive setting involves $80\%$ labelled nodes for training, this leads to a substantial increase of the training data size on three graphs, considering the transductive setting only uses a small portion of the labelled nodes to train the models. As a side effect, on Yelp graph with a 100,000 bump in the size of training nodes, a slight performance gain is observed from several deep methods, e.g., GAT, HAN, HGT and WIDEN. 
 
\begin{figure}[t]
\centering
\renewcommand{\arraystretch}{1.0}
\setlength\tabcolsep{2pt}
\begin{tabular}{ccc}
	\hspace{-0.3cm}\includegraphics[height=0.9in]{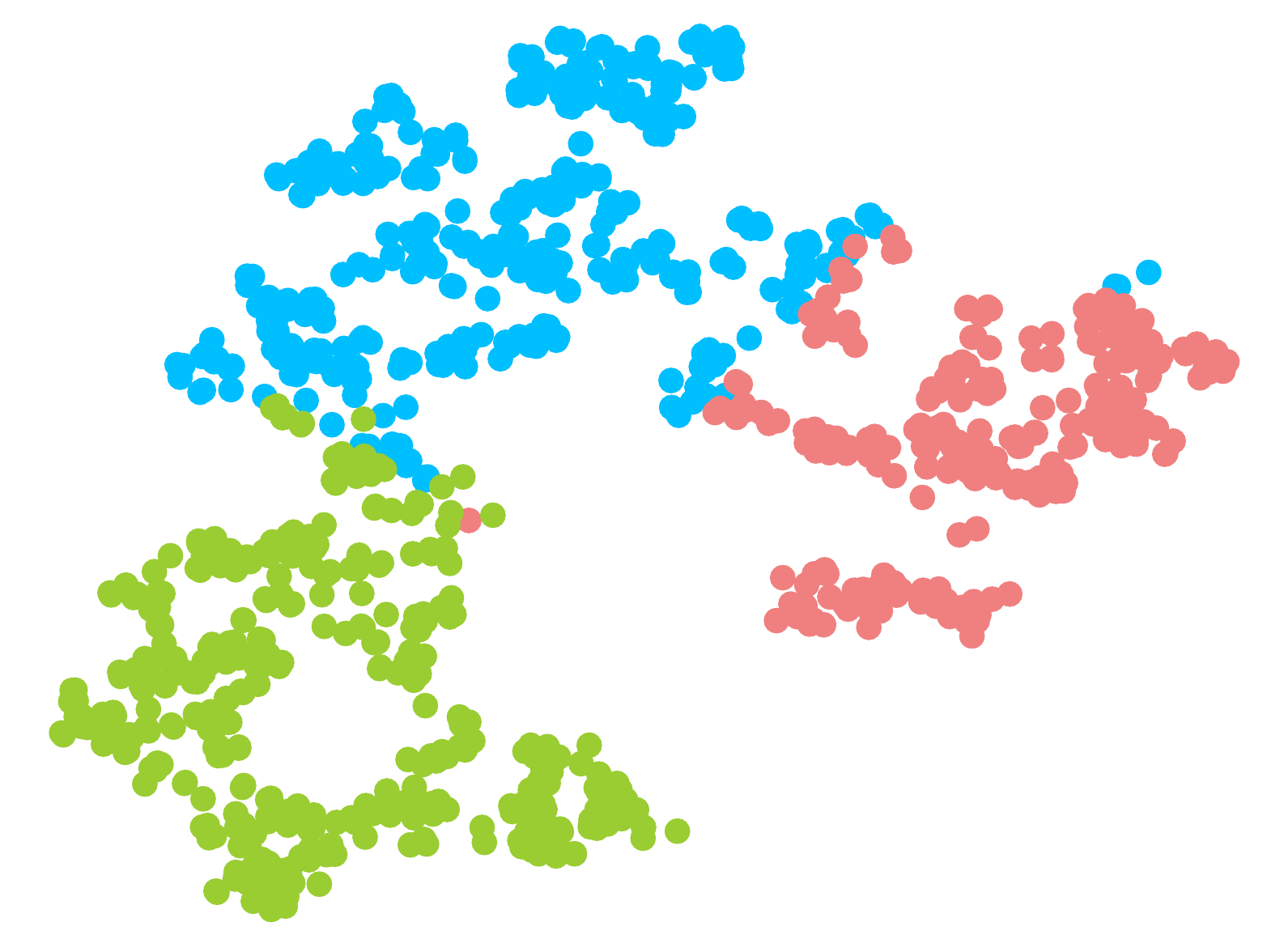}
	& \hspace{-0.2cm}\includegraphics[height=0.9in]{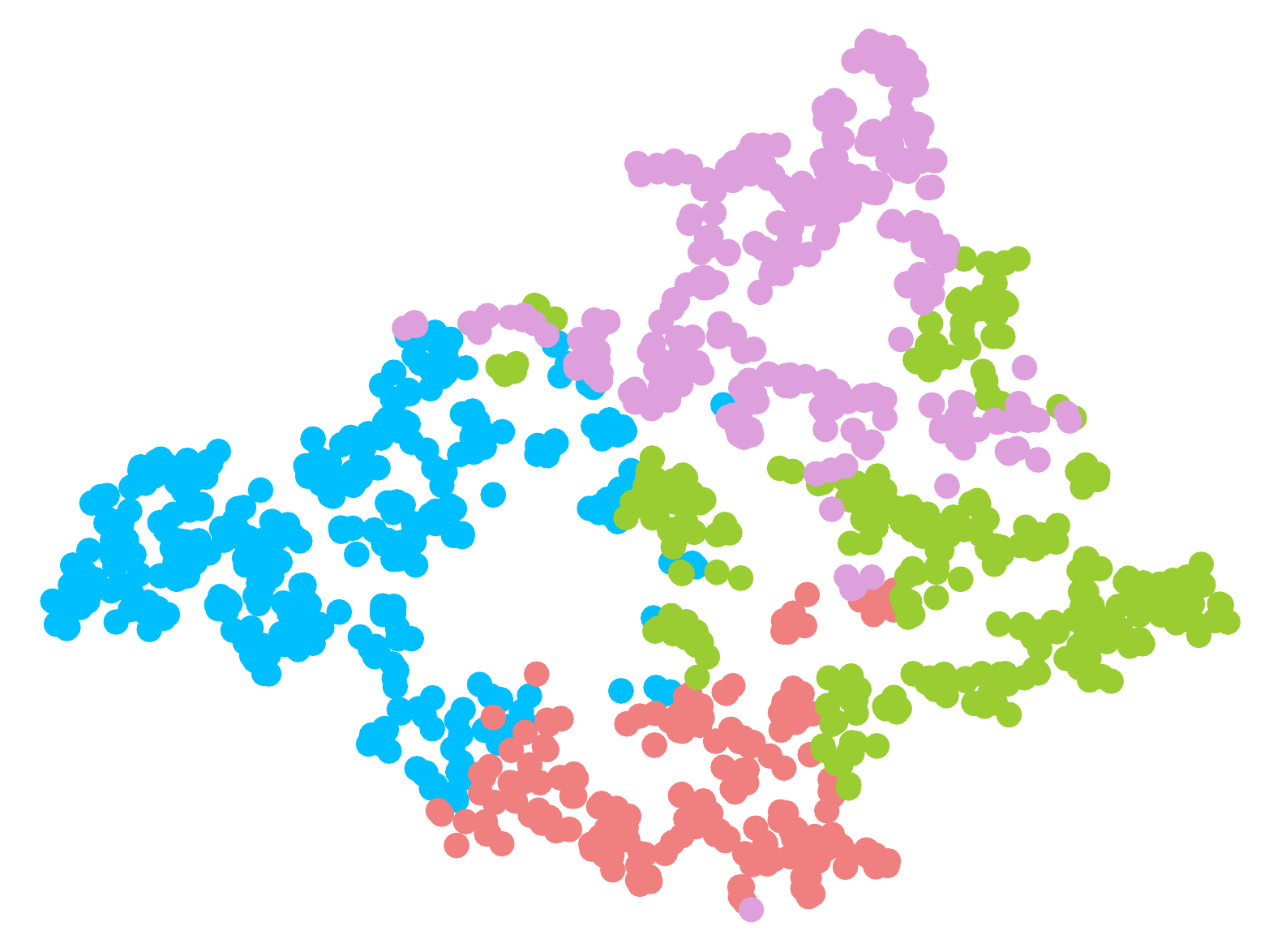}
	& \hspace{-0.2cm}\includegraphics[height=0.9in]{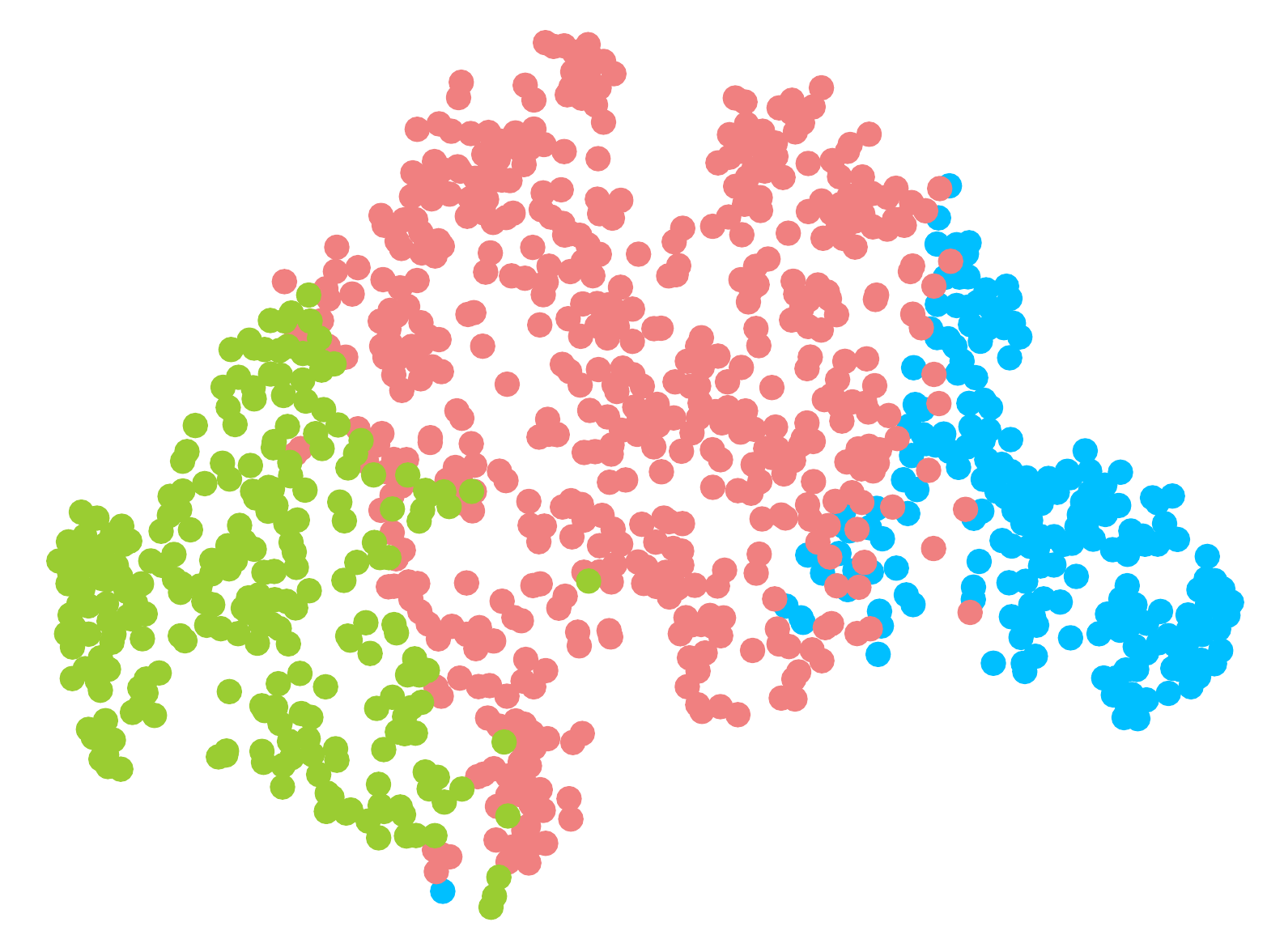}\\
	\footnotesize{(a) ACM}
	&\footnotesize{(b) DBLP}
	&\footnotesize{(b) Yelp}\\
\end{tabular}
\vspace{-0.4cm}
\caption{t-SNE visualization of inductively learned node embeddings on ACM, DBLP, and Yelp. Each node class in the dataset corresponds to one colour in the plot. Note that for Yelp, we randomly sampled 1,000 inductive nodes for better clarity.}
\label{figure:tSNE}
\vspace{-0.5cm}
\end{figure}

\textbf{Qualitative Results.} To qualitatively understand WIDEN's capability of performing inductive node representation learning on heterogeneous graphs, we organize a case study by visualizing the inductively learned node embeddings on all three datasets. Specifically, we translate the learned embeddings of inductive nodes into 2-dimensional vectors with t-SNE \cite{maaten2008visualizing}, which are then plotted in Figure \ref{figure:tSNE}. We use different colours to distinguish nodes from different classes. As can be seen from Figure \ref{figure:tSNE}, nodes from the same class tend to form their own cluster, and there are clear boundaries between two difference classes. Thus, even if all these nodes are new to WIDEN, it is still able to generate high-quality node representations owing to its comprehensive message passing architecture with the important edge information, and both wide and deep neighbors.

\subsection{Training Efficiency and Scalability (RQ3)}
Unlike most existing heterogeneous GNNs \cite{wang2019heterogeneous,hu2020heterogeneous}, one advantage of WIDEN is that it bypasses the sophisticated design for modelling relation-specific message passing processes. The message packaging scheme allows different semantics to be carried and passed into target nodes with a unified message passing architecture. In this section, we test the training efficiency and scalability of WIDEN given the importance of heterogeneous GNN's practicality in real-life applications. We discuss both aspects below.

\begin{figure*}[t!]
\centering
\begin{tabular}{cccc}
	\hspace{-0.4cm}\includegraphics[height=1.3in]{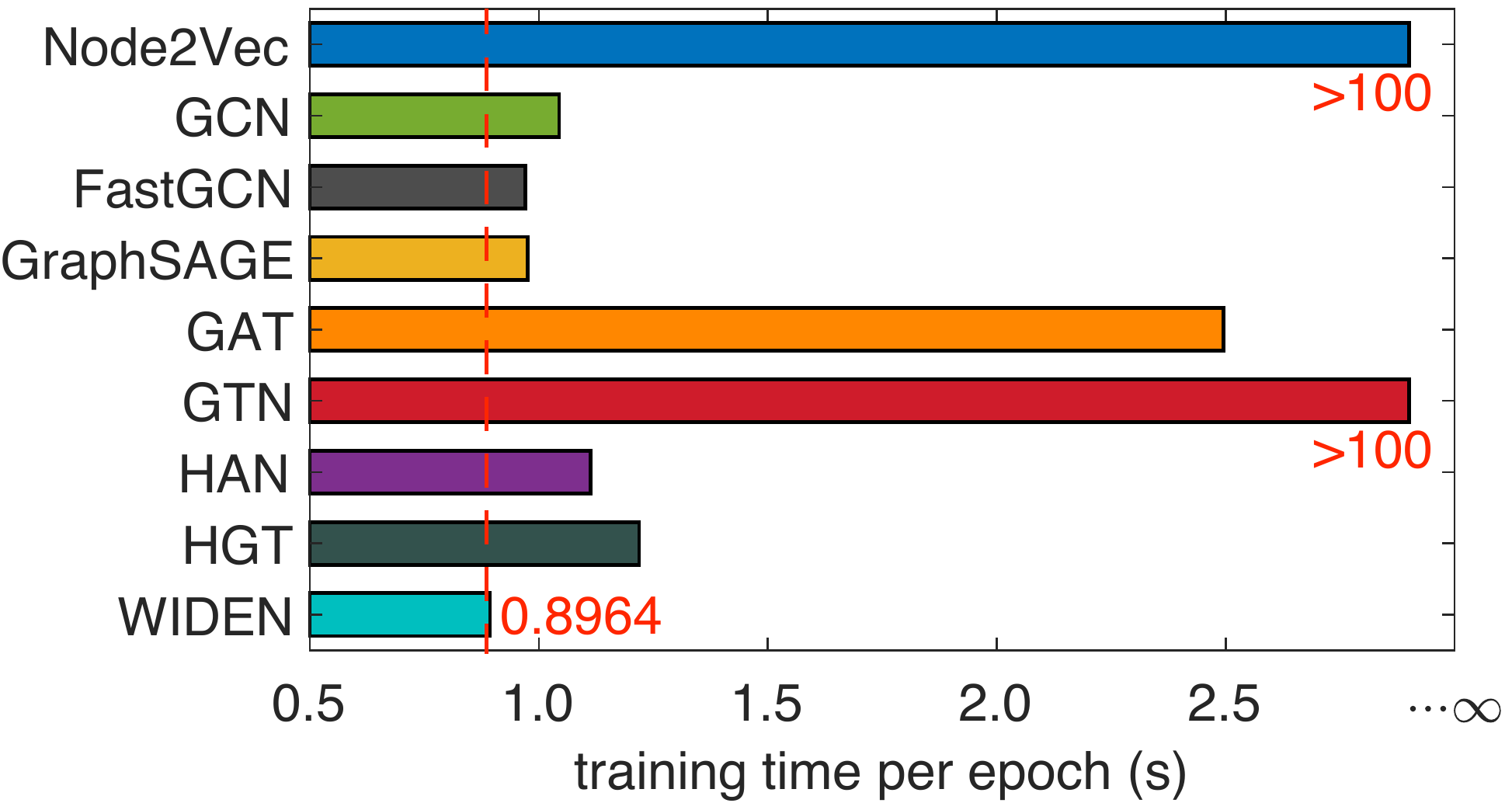}
	& \hspace{-0.4cm}\includegraphics[height=1.3in]{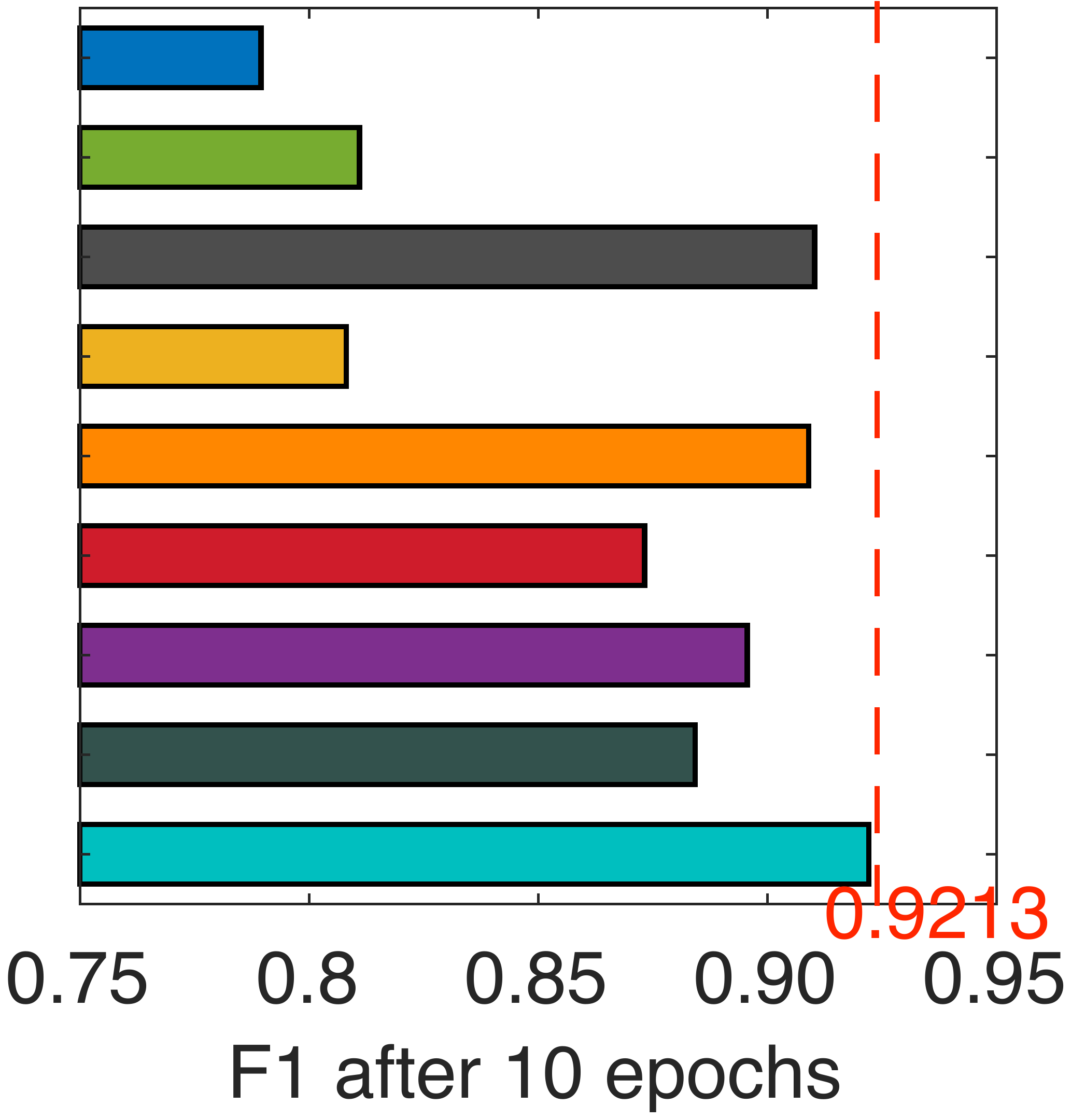}
	& \hspace{-0.4cm}\includegraphics[height=1.3in]{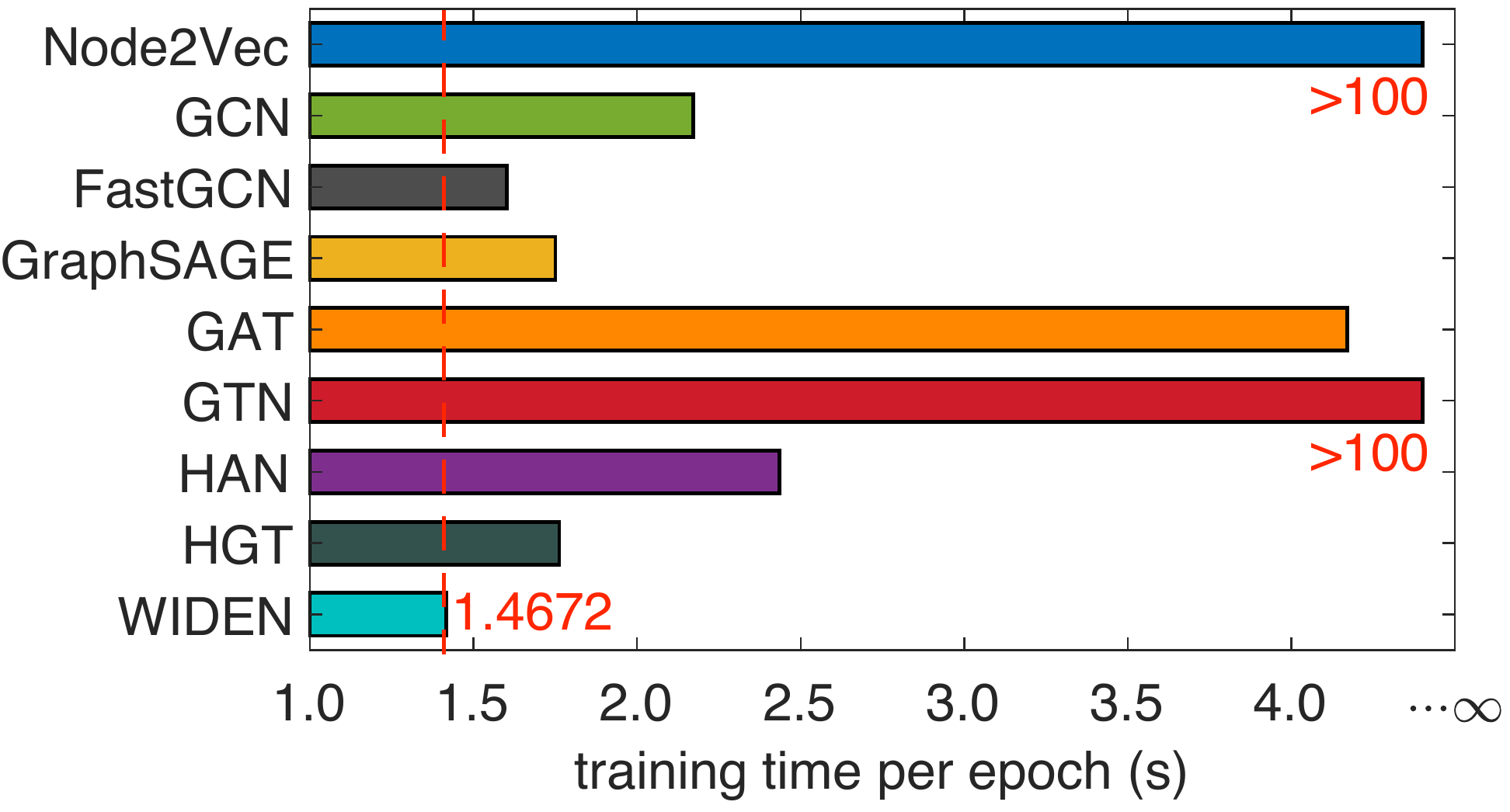}
	& \hspace{-0.4cm}\includegraphics[height=1.3in]{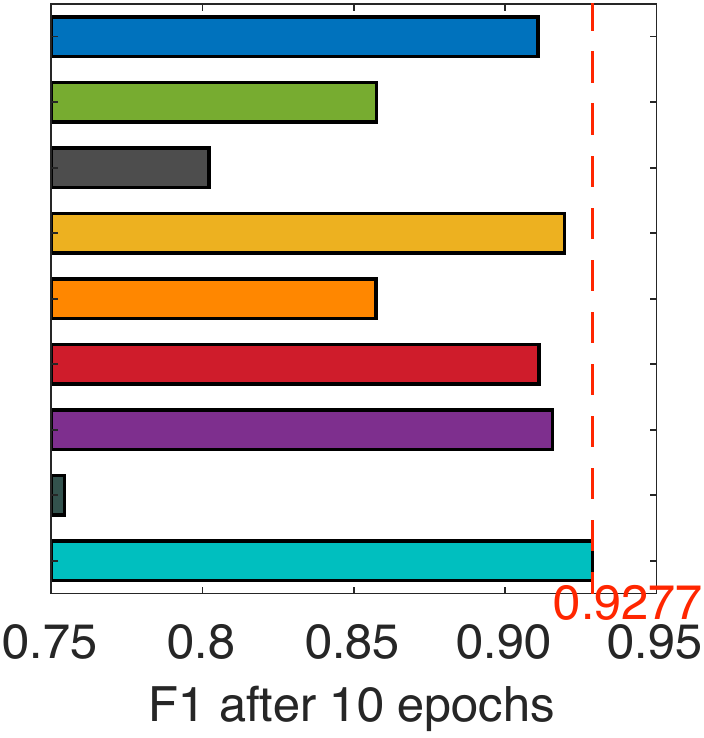}\\
	\multicolumn{2}{c}{\footnotesize{(a) Training efficiency comparison on ACM.}}
	&\multicolumn{2}{c}{\footnotesize{(b) Training efficiency comparison on DBLP.}}\\
\end{tabular}
\vspace{-0.4cm}
\caption{Training efficiency results. F1 scores after 10 training epochs are reported for reference, and the training time per epoch is averaged over all the training epochs taken by each method.}
\label{figure:training_efficiency}
\vspace{-0.5cm}
\end{figure*}

\begin{figure}[!t]
\vspace{-0.2cm}
\center
\includegraphics[width = 2.8in]{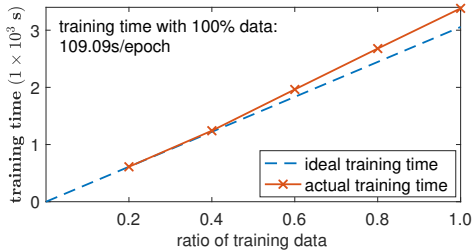}
\vspace{-0.4cm}
\caption{Training time of WIDEN on Yelp w.r.t varied data proportions.}
\label{Figure:train}
\vspace{-0.4cm}
\end{figure}

\textbf{Training Efficiency of WIDEN.} To make the training efficiency measurable, we examine all methods' training time consumption per epoch on ACM and DBLP\footnote{Hardware configuration: Intel Xeon Gold 6128 for CPU-enabled methods and Nvidia GeForce RTX 2080 Ti for GPU-enabled methods.}, and further evaluate their performance on (transductive) node classification after $10$ training epochs. As such, if it takes less time for a model to be trained over one epoch and the model yields better performance with the same amount of training epochs, the model is regarded highly efficient to train. As discussed in Section \ref{sec:settings}, on Yelp graph, most baseline methods can only be trained on one sampled sub-graph at a time, hence this part of the test is only conducted on the small-scale ACM and DBLP graphs. We demonstrate the results on training efficiency of all models in Figure \ref{figure:training_efficiency}. On both ACM and DBLP, WIDEN consumes as few as $0.8964$s and $0.9213$s to be trained over one epoch, which is faster than the relatively lightweight GraphSAGE and FastGCN (both need more than $1$s). The reasons behind are two-fold. First, WIDEN deploys the highly efficient self-attention modules to facilitate message passing between nodes. Second, the downsampling strategy directly reduces the amount of computations needed in each iteration, thus greatly cutting down the time consumption during training. Meanwhile, by outperforming all baselines in node classification on both graphs after only $10$ training epochs, WIDEN has demonstrated highly competitive training efficiency.

\textbf{Scalability of WIDEN.} We test the training efficiency and scalability of WIDEN by varying the proportions of the nodes in the graph. This is achieved by randomly sampling nodes from the full graph with a ratio of $\{0.2, 0.4, 0.6, 0.8, 1.0\}$ and then report the corresponding time cost for model training. The Yelp graph is used for scalability test since it is the largest graph in our experiments. The growth of training time along with the data size is shown in Figure \ref{Figure:train}. When the ratio of training data gradually extends from $0.2$ to $1.0$, the training time for SeqFM increases from $0.61\times10^{3}$ seconds to $3.38\times10^{3}$ seconds. It shows that the dependency of training time on the data scale is approximately linear. Hence, we conclude that WIDEN is scalable to even larger heterogeneous graphs. 

\subsection{Importance of Key Components (RQ4)}
To better understand the performance gain from the major components proposed in WIDEN, we conduct ablation tests on different degraded versions of WIDEN. Each variant removes one key component from the model, and the corresponding results on node classification are reported. Table \ref{table:ablation} summarizes node classification outcomes with different model architectures. In what follows, we introduce the variants and analyze their effect respectively.

\textbf{No Downsampling}. The adaptive downsampling strategy aims to selectively locate important node/edge information and improve the training efficiency with optimal performance. Removing the downsampling component will make the wide and deep node neighbor sets remain unchanged during the entire training process. As we can infer from Table \ref{table:ablation}, the resulted model performance is very similar on DBLP and shows a slight increase on ACM and Yelp. This is within our expectation as retaining all neighbor nodes during training will offer more information for message passing. As the same time, it indicates that our proposed downsampling strategy is fully able to ensure high-quality node representation learning.

\begin{table}[t]
\small
\vspace{-0.3cm}
\caption{Ablation test with different model architectures, where ``$\downarrow$" marks a severe (over $5\%$) performance drop.}
\vspace{-0.3cm}
\centering
\renewcommand{\arraystretch}{1.0}
\setlength\tabcolsep{0.5pt}
  \begin{tabular}{|c|c|c|c|}
    \hline
     Architecture & ACM & DBLP & Yelp\\
    \hline
    Default & \textbf{0.9269} & \textbf{0.9330} & \textbf{0.7179} \\
    No Downsampling & 0.9352 & 0.9323 & 0.7334 \\
    Removing Wide Neighbors & 0.9046 & 0.9023 & 0.7024\\
    Removing Deep Neighbors & 0.8976 & $\,\,\,$0.8126$\downarrow$ & $\,\,\,$0.6720$\downarrow$\\
    Removing Successive Self-Attention & 0.9035 & $\,\,\,$0.8832$\downarrow$ & 0.6913 \\
    Removing Relay Edges & 0.8885 & 0.8915 & 0.6947\\
    Random Downsampling for $\mathcal{W}(t)$ & 0.9192 & 0.9110 & 0.7111\\
    Random Downsampling for $\mathcal{D}(t)$ & $\,\,\,$0.8743$\downarrow$ & $\,\,\,$0.8537$\downarrow$ & 0.6867\\
     \hline
    \end{tabular}
\label{table:ablation}
\vspace{-0.6cm}
\end{table}

\textbf{Removing Wide or Deep Neighbors}. The second and third variant of WIDEN removes the message passing from wide and deep neighbors, respectively. Both variants suffer from inferior performance, especially when deep neighbors are removed from WIDEN since a severe (over $5\%$) performance drop is observed on both DBLP and Yelp graphs. Also, when wide neighbors are totally removed, WIDEN can still learn expressive node representations solely from its deep neighbor sets. This verifies our motivation of enriching the messages passed to the target node with sampled deep node sequences. 

\begin{figure*}[t!]
\centering
\begin{tabular}{cccc}
	\vspace{-0.1cm}\includegraphics[width=1.75in]{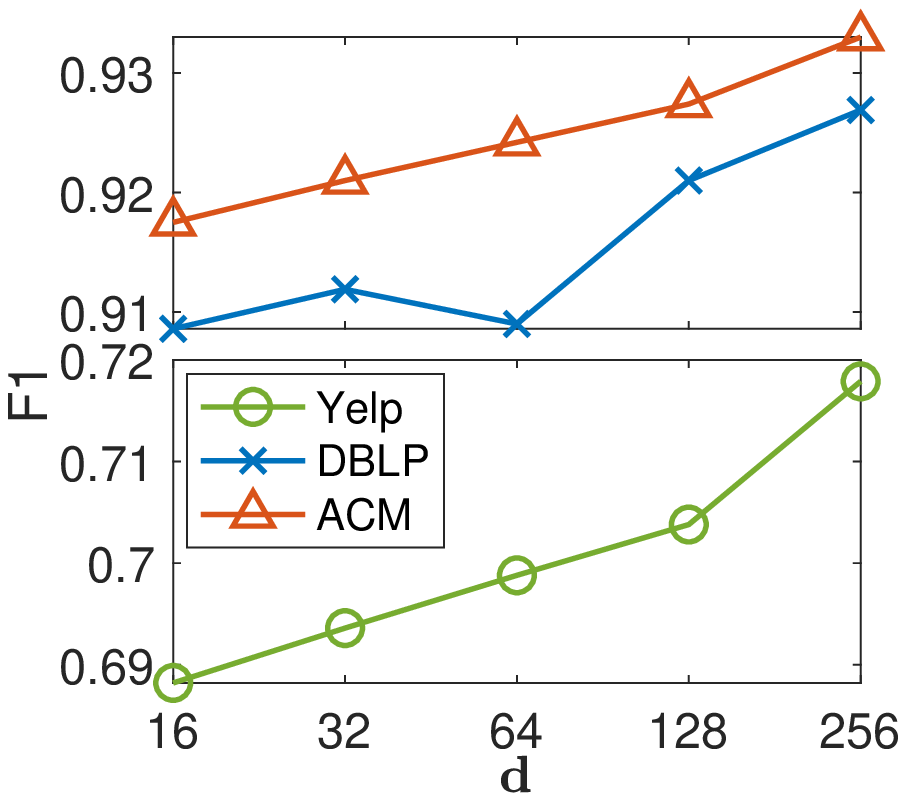}
	& \hspace{-0.6cm}\includegraphics[width=1.75in]{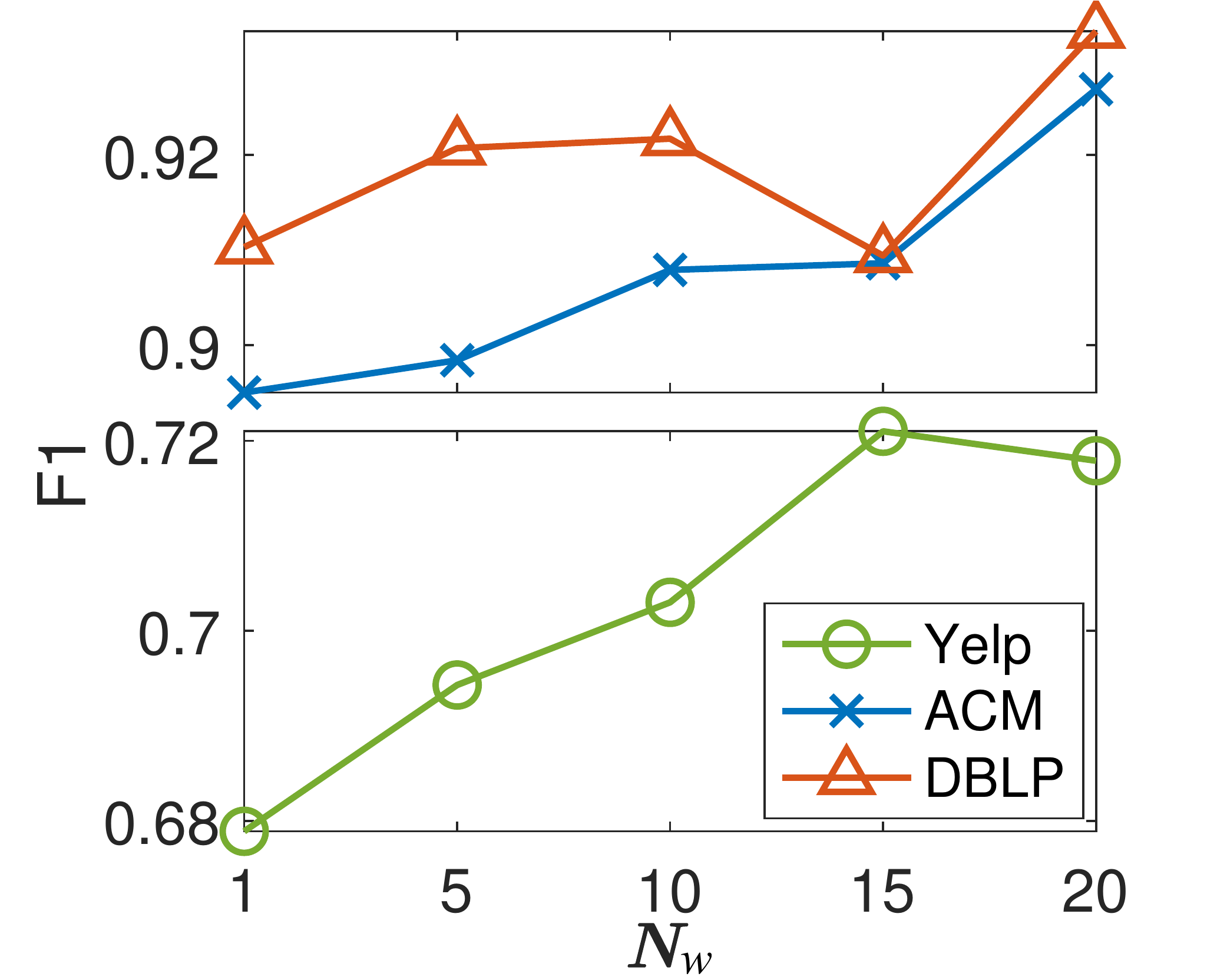}
	& \hspace{-0.6cm}\includegraphics[width=1.75in]{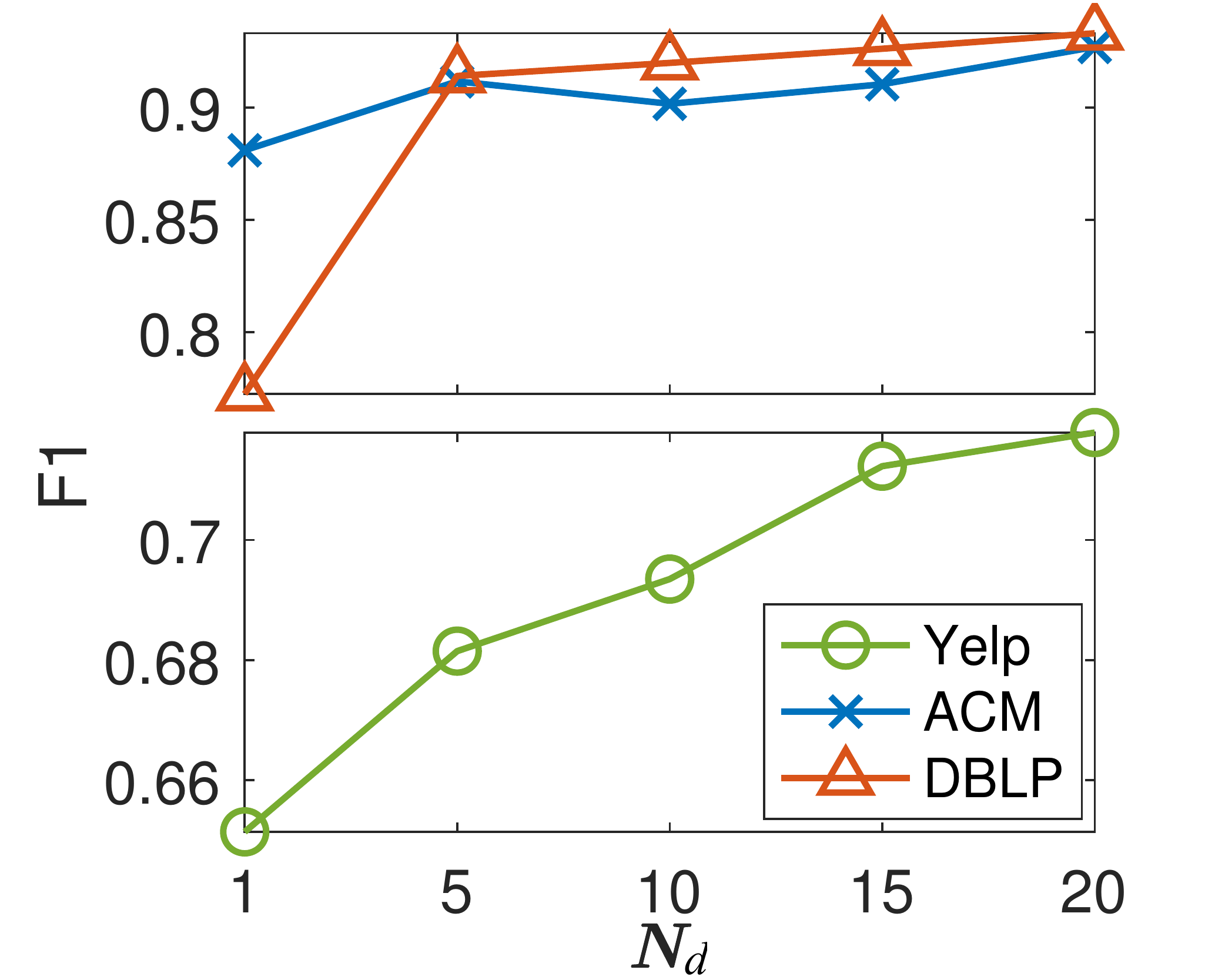}
	& \hspace{-0.6cm}\includegraphics[width=1.75in]{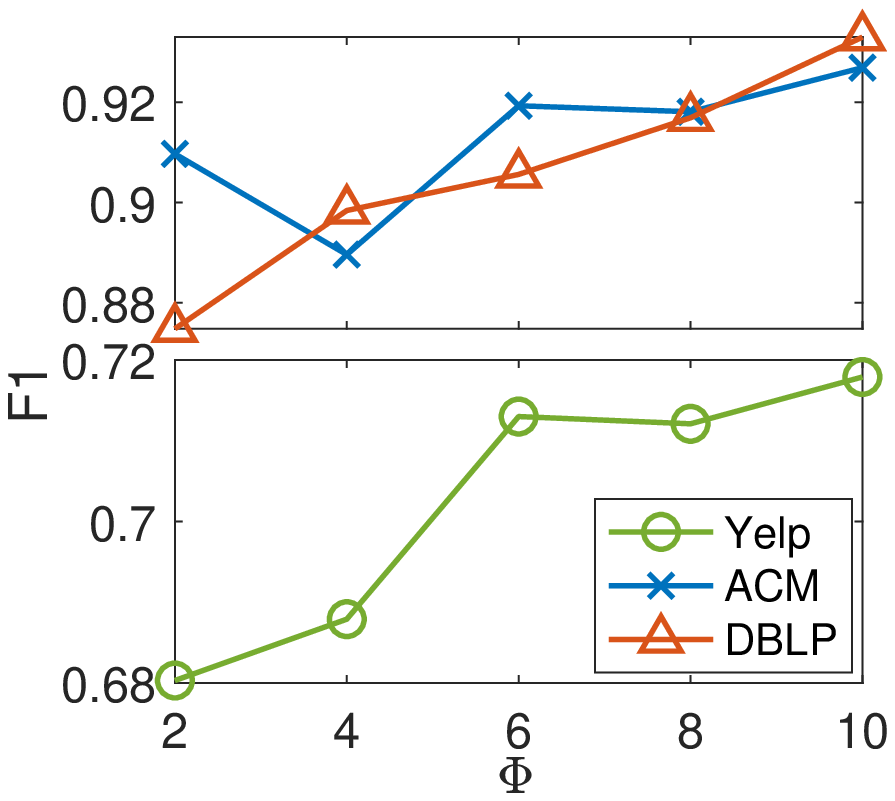}\\
	\footnotesize{(a) F1 Score of WIDEN w.r.t. $d$.}
	&\footnotesize{(b) F1 Score of WIDEN w.r.t. $N_w$.}
	&\footnotesize{(c) F1 Score of WIDEN w.r.t. $N_d$.}
	&\footnotesize{(d) F1 Score of WIDEN w.r.t. $\Phi$.}\\
\end{tabular}
	\vspace{-0.3cm}
\caption{Hyperparameter sensitivity analysis.}
\label{figure:paramsensitivity}
\vspace{-0.2cm}
\end{figure*}

\textbf{Removing Successive Self-Attention}. After removing the successive self-attention in Eq.(\ref{eq:deepMP1}), the deep message passing part is simplified into an attentive aggregation of all deep neighbor nodes w.r.t. the target node. This leads to worse model performance on all three datasets as removing this components limits WIDEN's ability to comprehensively capture the sequential dependencies among the message packs within the sampled deep walk sequences.

\textbf{Removing Relay Edges}. Without the relay edges, the deprecated message packs in the downsampling stage are directly discarded. As the relay edges help retain the semantics within the original deep walk sequences, deleting this component apparently makes the performance deteriorates. 

\textbf{Random Downsampling for $\mathcal{W}(t)$ or $\mathcal{D}(t)$}. Instead of using the computed attention scores for identifying irrelevant nodes, we randomly drop one node from either  $\mathcal{W}(t)$ or $\mathcal{D}(t)$ during the downsampling process. Note that the KL-divergence triggering mechanism is also removed in this variant. As can be seen from Table \ref{table:ablation}, randomly downsampling deep neighbors leads to more significant decrease on WIDEN's performance compared with randomly dropping wide neighbors. One reason is that, as $\mathcal{D}(t)$ contains higher-order connectivity information regarding the target node, using random downsampling instead of the attentive strategy is likely to drop out informative deep neighbors at an early stage, causing a substantial loss of useful information.

\subsection{Hyperparameter Sensitivity (RQ5)} \label{sec:hypersens}
We answer the forth research question by investigating the performance fluctuations of WIDEN with varied hyperparameters. Particularly, as mentioned in Section \ref{sec:settings}, we study our model's sensitivity to the latent dimension $d$, the number of wide neighbors $N_w$, the number of deep neighbors $N_d$, as well as the number of deep walk sequences for each target node $N_t$. For each test, based on the standard setting $\{d=256, N_w = 20, N_d = 20, N_t = 10\}$, we vary the value of one hyperparameter while keeping the others unchanged, and record the new F1 score achieved on transductive node classification. Figure \ref{figure:paramsensitivity} lays out the results with different parameter settings.

\textbf{Impact of $d$}.
The value of the embedding dimension $d$ is examined in $\{16,32,64,128,256\}$. As an important hyperparameter in deep neural networks, the latent dimension is apparently associated with the model's expressiveness. In general, WIDEN benefits from a relatively larger $d$ when learning node representations, hence setting $d = 256$ yields the highest F1 scores for node classification tasks. It is worth mentioning that on Yelp graph, even with $d=16$, WIDEN still outperforms all the baselines, which further proves the effectiveness of our proposed model on large-scale heterogeneous graphs.

\textbf{Impact of $N_w$}.
We study the impact of the number of initial neighbor nodes in each target node's wide neighbor set with $N_w \in \{1,5,10,15,20\}$. For both ACM and DBLP graphs, a total of 20 initial neighbors in the wide neighbor set leads to the best performance, while inferior performance is observed when $N_w$ is reduced. Interestingly, on Yelp graph, when $N_w=15$, WIDEN achieves slightly better performance than having $N_w = 20$. This is possibly because setting $N_w$ to $15$ also helps remove neighbor nodes that might contain noisy information. 

\textbf{Impact of $N_d$}.
As can be concluded from Figure \ref{figure:paramsensitivity}, WIDEN's performance is positively associated with the number of deep neighbor nodes $N_d$, which are examined in $\{1, 5, 10, 15, 20\}$. Compared with smaller graphs ACM and DBLP, a larger $N_d$ brings more significant performance increase on Yelp graph. This validates that on large and sparse heterogeneous graphs, passing information from remotely connected nodes is actually beneficial for node representation learning.

\textbf{Impact of $\Phi$}.
The impact of different numbers of deep walk sequences for a target node is investigated via $\Phi \in \{2,4,6,8,10\}$. Overall, the best prediction performance of SeqFM is reached when $\Phi = 10$ and more sampled deep walk sequences generally yield better results. We do notice that the performance gain starts to become less significant when $\Phi$ reaches a certain scale ($6$ in our case), indicating chances for a further boost in WIDEN's efficiency with small trade-offs in performance.

\section{Related Work}
In this section, we provide a summary of works related to our paper. We will start with classic (i.e., deep learning-free) graph representation learning methods, and then extend to state-of-the-art graph neural networks.

\subsection{Deep Learning-free Graph Representation Learning Methods}
Graph representation learning, i.e., graph embedding is the task of embedding a graph into a low-dimensional space while preserving the network structure and property to support downstream tasks \cite{wang2019heterogeneous}. Most existing graph embedding tasks focus on learning representations for nodes, where methods based on random walk \cite{grover2016node2vec,perozzi2014deepwalk,qiu2018network} are initially proposed to learn node embeddings on homogeneous graphs. Originated from the skip-gram learning algorithm \cite{mikolov2013efficient}, random walk-based methods samples node sequences from a graph, then learns node embeddings by optimizing the co-occurrence probability of nodes within the same sequence. Meanwhile, LINE \cite{tang2015line} introduces two hypothesis where two nodes are highly similar in the embedding space if they have direction connections (i.e., first-order proximity) and/or similar neighborhood compositions (i.e., second-order proximity), based on which two effective loss functions are designed for graph embedding.

To cope with the heterogeneity of nodes and edges, \cite{tang2015pte} proposes PTE model that extends the notion of second-order proximity \cite{tang2015line} to heterogeneous graphs by separating a graph into multiple bipartite/homogeneous graphs. In a similar spirit, Chen et al. proposes to learn relation-specific projection matrices \cite{chen2018pme} in the divided bipartite/homogeneous graphs in order to learn more expressive node embeddings. Among heterogeneous graph embedding methods, a large body of recent works \cite{fu2017hin2vec,cen2019representation,he2019hetespaceywalk,park2019task} make use of the heterogeneous contexts within meta paths \cite{sun2011pathsim} to facilitate node representation learning for downstream tasks. Essentially, a meta path is a sequence of heterogeneous nodes linked via various types of relations \cite{dong2020heterogeneous}. For example, \cite{fu2017hin2vec} directly considers meta paths as contextual information when learning embeddings for nodes, while He et al. \cite{he2019hetespaceywalk} expand the meta path-based random walk by proposing a heterogeneous personalized spacey random walk algorithm based on Markov chains.

Most deep learning-free graph embedding methods focus on improving the sampling strategies to acquire richer contextual information (e.g., generating high-quality random walk sequences \cite{he2019hetespaceywalk,grover2016node2vec}) for learning node embeddings, or enhancing the optimization paradigm to maximize the structural and property information preserved in the learned embedding (e.g., defining contrastive losses \cite{garcia2017learning,tang2015line}). Though proven effective, these methods are subject to limited expressiveness for representation learning. In what follows, we will discuss the deep graph neural networks which have been highly advantageous in a considerable amount of graph embedding tasks.

\subsection{Graph Neural Networks}
Recent years have witnessed the prominence of graph neural networks (GNNs) in modelling structural \cite{bruna2014spectral} and relational data \cite{schlichtkrull2018modeling}. Different from aforementioned shallow embedding methods, GNNs are often empowered by more complex encoders, usually a deep neural network, enabling the natural modeling of both structures and vertex attributes \cite{dong2020heterogeneous,chen2018fastgcn,velickovic2019deep,chen2019exploiting,wang2019origin,sun2020disease}. In general, the inner mechanism of GNNs can be viewed as a message passing process, like propagating node or edge information via the graph structure in MPNN \cite{gilmer2017neural} or treating the neighbors of the target node as a receptive field for iterative feature aggregation in GCN \cite{kipf2017GCN}, GraphSAGE \cite{hamilton2017inductive} and GAT \cite{velivckovic2018graph}.

As an early variant of GNNs for modelling heterogeneous graphs, the relational GCN \cite{schlichtkrull2018modeling} learns a distinct linear projection weight for each edge type.
%Another notable effort is the Decagon model [Zitnik et al., 2018], which applies specific graph convolutional filters over each type of relations in a multimodal graph with proteins, drugs, polypharmacy side effects as vertices. Note that the relation-specific parameters in Decagon are shared for differ- ent vertices.
Similarly, to deal with heterogeneous graph structures and node attributes, \cite{zhang2019heterogeneous} uses type-specific RNNs to encode features for each type of neighbor vertices, followed by a subsequent RNN to aggregate different types of neighbor representations. 
More recently, the heterogeneous graph transformer (HGT) \cite{hu2020heterogeneous} uses each edge's meta relation to parameterize the transformer network \cite{vaswani2017attention}, so that both the common and specific patterns of different relationships are effectively captured. 

Concurrently, an increasing amount of efforts have been devoted to enhancing heterogeneous GNNs with meta paths and attention mechanisms \cite{chen2020social,shi2018heterogeneous,wang2019heterogeneous,yun2019graph,hu2020heterogeneous,zhang2020gcn}. For example, \cite{hu2018leveraging,shi2018heterogeneous} utilize the rich contexts within heterogeneous meta paths to enhance the representation learning capability of deep neural networks for recommendation tasks. Afterwards, the heterogeneous graph attention network (HAN) \cite{yun2019graph} defines different attentive aggregators and weight matrices for each meta path when embedding the target node. To bypass the inflexibility of manually defined meta paths, Yun et al. propose the graph transformer network (GTN) \cite{yun2019graph}, which learns a soft selection of edge types and composite relations for generating useful multi-hop meta paths to augment the input graph. After that, plain GCNs are directly adopted in GTN for learning contextualized node embeddings.

\subsection{Conclusion}
This paper presents our effort in uniting heterogeneity, inductiveness, and efficiency for graph representation learning. To address those three practical problems, we propose WIDEN, which establishes a heterogeneous message packaging and passing paradigm for both wide and deep neighbor nodes that fully supports inductive representation learning. Furthermore, we design an active downsampling approach to facilitate efficient training while preserving the expressiveness of the learned node embeddings. Experimental results on real-world datasets have demonstrated the efficacy of WIDEN in resolving the three important challenges when learning node representations on graphs.

\ifCLASSOPTIONcaptionsoff
  \newpage
\fi



\begin{thebibliography}{10}
\providecommand{\url}[1]{#1}
\csname url@samestyle\endcsname
\providecommand{\newblock}{\relax}
\providecommand{\bibinfo}[2]{#2}
\providecommand{\BIBentrySTDinterwordspacing}{\spaceskip=0pt\relax}
\providecommand{\BIBentryALTinterwordstretchfactor}{4}
\providecommand{\BIBentryALTinterwordspacing}{\spaceskip=\fontdimen2\font plus
\BIBentryALTinterwordstretchfactor\fontdimen3\font minus
  \fontdimen4\font\relax}
\providecommand{\BIBforeignlanguage}[2]{{%
\expandafter\ifx\csname l@#1\endcsname\relax
\typeout{** WARNING: IEEEtran.bst: No hyphenation pattern has been}%
\typeout{** loaded for the language `#1'. Using the pattern for}%
\typeout{** the default language instead.}%
\else
\language=\csname l@#1\endcsname
\fi
#2}}
\providecommand{\BIBdecl}{\relax}
\BIBdecl

\bibitem{wang2019neural}
X.~Wang, X.~He, M.~Wang, F.~Feng, and T.-S. Chua, ``Neural graph collaborative
  filtering,'' in \emph{SIGIR}, 2019.

\bibitem{butler2018machine}
K.~T. Butler, D.~W. Davies, H.~Cartwright, O.~Isayev, and A.~Walsh, ``Machine
  learning for molecular and materials science,'' \emph{Nature}, vol. 559, no.
  7715, pp. 547--555, 2018.

\bibitem{kipf2017GCN}
T.~N. Kipf and M.~Welling, ``Semi-supervised classification with graph
  convolutional networks,'' \emph{ICLR}, 2017.

\bibitem{velivckovic2018graph}
P.~Veli{\v{c}}kovi{\'c}, G.~Cucurull, A.~Casanova, A.~Romero, P.~Li{\`o}, and
  Y.~Bengio, ``Graph attention networks,'' in \emph{ICLR}, 2018.

\bibitem{yun2019graph}
S.~Yun, M.~Jeong, R.~Kim, J.~Kang, and H.~J. Kim, ``Graph transformer
  networks,'' in \emph{NIPS}, 2019, pp. 11\,983--11\,993.

\bibitem{wang2019heterogeneous}
X.~Wang, H.~Ji, C.~Shi, B.~Wang, Y.~Ye, P.~Cui, and P.~S. Yu, ``Heterogeneous
  graph attention network,'' in \emph{WWW}, 2019, pp. 2022--2032.

\bibitem{hu2020heterogeneous}
Z.~Hu, Y.~Dong, K.~Wang, and Y.~Sun, ``Heterogeneous graph transformer,'' in
  \emph{The Web Conference}, 2020, pp. 2704--2710.

\bibitem{wang2019kgat}
X.~Wang, X.~He, Y.~Cao, M.~Liu, and T.-S. Chua, ``Kgat: Knowledge graph
  attention network for recommendation,'' in \emph{SIGKDD}, 2019, pp. 950--958.

\bibitem{sun2011pathsim}
Y.~Sun, J.~Han, X.~Yan, P.~S. Yu, and T.~Wu, ``Pathsim: Meta path-based top-k
  similarity search in heterogeneous information networks,'' \emph{VLDB
  Endowment}, vol.~4, no.~11, pp. 992--1003, 2011.

\bibitem{zhang2019heterogeneous}
C.~Zhang, D.~Song, C.~Huang, A.~Swami, and N.~V. Chawla, ``Heterogeneous graph
  neural network,'' in \emph{SIGKDD}, 2019, pp. 793--803.

\bibitem{park2020meta}
S.-w. Park, B.~J. Bae, J.~Yeo, and S.-w. Hwang, ``Meta-path free
  semi-supervised learning for heterogeneous networks,'' \emph{arXiv preprint
  arXiv:2010.08924}, 2020.

\bibitem{schlichtkrull2018modeling}
M.~Schlichtkrull, T.~N. Kipf, P.~Bloem, R.~Van Den~Berg, I.~Titov, and
  M.~Welling, ``Modeling relational data with graph convolutional networks,''
  in \emph{ESWC}, 2018, pp. 593--607.

\bibitem{kang2018self}
W.-C. Kang and J.~McAuley, ``Self-attentive sequential recommendation,''
  \emph{ICDM}, pp. 197--206, 2018.

\bibitem{nilizadeh2019think}
S.~Nilizadeh, H.~Aghakhani, E.~Gustafson, C.~Kruegel, and G.~Vigna, ``Think
  outside the dataset: Finding fraudulent reviews using cross-dataset
  analysis,'' in \emph{WWW}, 2019, pp. 3108--3115.

\bibitem{hamilton2017inductive}
W.~Hamilton, Z.~Ying, and J.~Leskovec, ``Inductive representation learning on
  large graphs,'' in \emph{NIPS}, 2017, pp. 1024--1034.

\bibitem{ying2018graph}
R.~Ying, R.~He, K.~Chen, P.~Eksombatchai, W.~L. Hamilton, and J.~Leskovec,
  ``Graph convolutional neural networks for web-scale recommender systems,'' in
  \emph{SIGKDD}, 2018.

\bibitem{chen2018fastgcn}
J.~Chen, T.~Ma, and C.~Xiao, ``Fastgcn: fast learning with graph convolutional
  networks via importance sampling,'' \emph{ICLR}, 2018.

\bibitem{gao2018bine}
M.~Gao, L.~Chen, X.~He, and A.~Zhou, ``Bine: Bipartite network embedding,'' in
  \emph{SIGIR}, 2018, pp. 715--724.

\bibitem{vaswani2017attention}
A.~Vaswani, N.~Shazeer, N.~Parmar, J.~Uszkoreit, L.~Jones, A.~N. Gomez,
  {\L}.~Kaiser, and I.~Polosukhin, ``Attention is all you need,'' in
  \emph{NIPS}, 2017.

\bibitem{mikolov2013efficient}
T.~Mikolov, K.~Chen, G.~Corrado, and J.~Dean, ``Efficient estimation of word
  representations in vector space,'' \emph{ICLR}, 2013.

\bibitem{perozzi2014deepwalk}
B.~Perozzi, R.~Al-Rfou, and S.~Skiena, ``Deepwalk: Online learning of social
  representations,'' in \emph{SIGKDD}, 2014, pp. 701--710.

\bibitem{grover2016node2vec}
A.~Grover and J.~Leskovec, ``node2vec: Scalable feature learning for
  networks,'' in \emph{SIGKDD}, 2016, pp. 855--864.

\bibitem{li2017deepcas}
C.~Li, J.~Ma, X.~Guo, and Q.~Mei, ``Deepcas: An end-to-end predictor of
  information cascades,'' in \emph{WWW}, 2017, pp. 577--586.

\bibitem{yang2019multi}
C.~Yang, J.~Tang, M.~Sun, G.~Cui, and Z.~Liu, ``Multi-scale information
  diffusion prediction with reinforced recurrent networks.'' in \emph{IJCAI},
  2019, pp. 4033--4039.

\bibitem{karypis1998fast}
G.~Karypis and V.~Kumar, ``A fast and high quality multilevel scheme for
  partitioning irregular graphs,'' \emph{SIAM Journal on Scientific Computing},
  vol.~20, no.~1, pp. 359--392, 1998.

\bibitem{maaten2008visualizing}
L.~v.~d. Maaten and G.~Hinton, ``Visualizing data using t-sne,'' \emph{JMLR},
  vol.~9, no. Nov, pp. 2579--2605, 2008.

\bibitem{qiu2018network}
J.~Qiu, Y.~Dong, H.~Ma, J.~Li, K.~Wang, and J.~Tang, ``Network embedding as
  matrix factorization: Unifying deepwalk, line, pte, and node2vec,'' in
  \emph{WSDM}, 2018, pp. 459--467.

\bibitem{tang2015line}
J.~Tang, M.~Qu, M.~Wang, M.~Zhang, J.~Yan, and Q.~Mei, ``Line: Large-scale
  information network embedding,'' in \emph{WWW}, 2015, pp. 1067--1077.

\bibitem{tang2015pte}
J.~Tang, M.~Qu, and Q.~Mei, ``Pte: Predictive text embedding through
  large-scale heterogeneous text networks,'' in \emph{SIGKDD}, 2015, pp.
  1165--1174.

\bibitem{chen2018pme}
H.~Chen, H.~Yin, W.~Wang, H.~Wang, Q.~V.~H. Nguyen, and X.~Li, ``Pme: projected
  metric embedding on heterogeneous networks for link prediction,'' in
  \emph{SIGKDD}, 2018.

\bibitem{fu2017hin2vec}
T.-y. Fu, W.-C. Lee, and Z.~Lei, ``Hin2vec: Explore meta-paths in heterogeneous
  information networks for representation learning,'' in \emph{CIKM}, 2017, pp.
  1797--1806.

\bibitem{cen2019representation}
Y.~Cen, X.~Zou, J.~Zhang, H.~Yang, J.~Zhou, and J.~Tang, ``Representation
  learning for attributed multiplex heterogeneous network,'' in \emph{SIGKDD},
  2019, pp. 1358--1368.

\bibitem{he2019hetespaceywalk}
Y.~He, Y.~Song, J.~Li, C.~Ji, J.~Peng, and H.~Peng, ``Hetespaceywalk: a
  heterogeneous spacey random walk for heterogeneous information network
  embedding,'' in \emph{CIKM}, 2019, pp. 639--648.

\bibitem{park2019task}
C.~Park, D.~Kim, Q.~Zhu, J.~Han, and H.~Yu, ``Task-guided pair embedding in
  heterogeneous network,'' in \emph{CIKM}, 2019, pp. 489--498.

\bibitem{dong2020heterogeneous}
Y.~Dong, Z.~Hu, K.~Wang, Y.~Sun, and J.~Tang, ``Heterogeneous network
  representation learning,'' in \emph{IJCAI}, 2020, pp. 4861--4867.

\bibitem{garcia2017learning}
A.~Garcia~Duran and M.~Niepert, ``Learning graph representations with embedding
  propagation,'' \emph{NIPS}, pp. 5119--5130, 2017.

\bibitem{bruna2014spectral}
J.~Bruna, W.~Zaremba, A.~Szlam, and Y.~Lecun, ``Spectral networks and locally
  connected networks on graphs,'' in \emph{ICLR}, 2014.

\bibitem{velickovic2019deep}
P.~Velickovic, W.~Fedus, W.~L. Hamilton, P.~Li{\`o}, Y.~Bengio, and R.~D.
  Hjelm, ``Deep graph infomax,'' in \emph{ICLR}, 2019.

\bibitem{chen2019exploiting}
H.~Chen, H.~Yin, T.~Chen, Q.~V.~H. Nguyen, W.-C. Peng, and X.~Li, ``Exploiting
  centrality information with graph convolutions for network representation
  learning,'' in \emph{ICDE}, 2019, pp. 590--601.

\bibitem{gilmer2017neural}
J.~Gilmer, S.~S. Schoenholz, P.~F. Riley, O.~Vinyals, and G.~E. Dahl, ``Neural
  message passing for quantum chemistry,'' \emph{ICML}, 2017.

\bibitem{chen2020social}
H.~Chen, H.~Yin, T.~Chen, W.~Wang, X.~Li, and X.~Hu, ``Social boosted
  recommendation with folded bipartite network embedding,'' \emph{TKDE}, 2020.

\bibitem{shi2018heterogeneous}
C.~Shi, B.~Hu, W.~X. Zhao, and S.~Y. Philip, ``Heterogeneous information
  network embedding for recommendation,'' \emph{TKDE}, vol.~31, no.~2, pp.
  357--370, 2018.

\bibitem{zhang2020gcn}
S.~Zhang, H.~Yin, T.~Chen, Q.~V.~N. Hung, Z.~Huang, and L.~Cui, ``Gcn-based
  user representation learning for unifying robust recommendation and fraudster
  detection,'' in \emph{SIGIR}, 2020, pp. 689--698.

\bibitem{hu2018leveraging}
B.~Hu, C.~Shi, W.~X. Zhao, and P.~S. Yu, ``Leveraging meta-path based context
  for top-n recommendation with a neural co-attention model,'' in
  \emph{SIGKDD}, 2018.
  
\bibitem{chen2019air}
T.~Chen, H.~Yin, H.~Chen, Q.~V.~H. Nguyen, W.~C. Peng, and X.~Li, ``Air: attentional intention-aware recommender systems,'' in
  \emph{ICDE}, 2019.
  
\bibitem{wang2019origin}
Y.~Wang, H.~Yin, H.~Chen, T.~Wo, J.~Xu, and K.~Zheng, ``Origin-destination matrix prediction via graph convolution: a new perspective of passenger demand modeling,'' in
  \emph{SIGKDD}, 2019.

\bibitem{chen2020sequence}
T.~Chen, H.~Yin, Q.~V.~H. Nguyen, W.~C. Peng, X.~Li and X.~Zhou, ``Sequence-aware factorization machines for predictive analytics,'' in
  \emph{ICDE}, 2020.
  
\bibitem{sun2020disease}
Z.~Sun, H.~Yin, H.~Chen, T.~Chen, L.~Cui and F.~Yang, ``Disease prediction via graph neural networks,'' in
  \emph{JBHI}, 2020.

\end{thebibliography}
\end{document}